\documentclass[aps,prd,twocolumn,groupedaddress]{revtex4}
\usepackage{graphicx}
\begin{document}

\title{Detailed analysis of the gluonic excitation 
in the three-quark system in lattice QCD}

\author{Toru~T.~Takahashi}
%\thanks{}
\affiliation{Yukawa Institute for Theoretical Physics, Kyoto University, 
Kitashirakawa-Oiwake, Sakyo, Kyoto 606-8502, Japan}

\author{Hideo~Suganuma}
%\thanks{}
\affiliation{Faculty of Science, Tokyo Institute of Technology,
Ohokayama 2-12-1, Meguro, Tokyo 152-8551, Japan}

\date{\today}

\begin{abstract}

We study the excited-state potential and the gluonic excitation in the static three-quark (3Q) system
using SU(3) lattice QCD with $16^3\times 32$ at $\beta$=5.8 and 6.0 at the quenched level. 
For about 100 different patterns of spatially-fixed 3Q systems, we accurately extract the excited-state potential 
$V_{\rm 3Q}^{\rm e.s.}$ together with the ground-state potential $V_{\rm 3Q}^{\rm g.s.}$ 
by diagonalizing the QCD Hamiltonian in the presence of three quarks. 
The gluonic excitation energy $\Delta E_{\rm 3Q} \equiv V_{\rm 3Q}^{\rm e.s.}-V_{\rm 3Q}^{\rm g.s.}$ 
is found to be about 1 GeV at the typical hadronic scale. 
This large gluonic-excitation energy is conjectured to give a physical reason of  
the success of the quark model for low-lying hadrons even without explicit gluonic modes.
We investigate the functional form of $\Delta E_{\rm 3Q}$ in terms of the 3Q location.
The lattice data of $\Delta E_{\rm 3Q}$ are relatively well reproduced by the ``inverse Mercedes Ansatz'' 
with the ``modified Y-type flux-tube length'', 
which indicates that the gluonic-excitation mode is realized as a complicated bulk excitation of the whole 3Q system. 
\end{abstract}

\maketitle

\section{Introduction}\label{sec1}

It is widely accepted that quantum chromodynamics (QCD) \cite{QCD6673}
is the fundamental theory of the strong interaction for hadrons and nuclei. 
Nevertheless, it still remains as a difficult problem 
to derive low-energy physical quantities directly from QCD in an analytic manner. 
The perturbative QCD calculation, which successfully describes the high-energy process, 
cannot be applied  at the hadronic scale, 
since the QCD coupling constant becomes large in the infrared region.
Furthermore, the strong-coupling nature of QCD leads to a highly nontrival vacuum with 
rich nonperturbative phenomena such as color confinement and spontaneous chiral symmetry breaking.

In this decade, the lattice-QCD Monte Carlo calculation has been recognized as 
a reliable nonperturbative method for the quantitative analysis of QCD,
and the nonperturbative analysis based on QCD has become 
one of the central and important issues in the hadron physics.
For instance, lattice QCD calculations successfully  
reproduce hadron mass spectra~\cite{CPPACS01} and also predict the QCD phase transition 
at finite temperatures and densities~\cite{K03}. 
On the other hand, the underlying structure of hadrons is not yet well investigated using lattice QCD.
In fact, the application of lattice QCD is just beginning 
for the research of the hadron structure  
and related excitation modes in terms of quarks and gluons. 

The investigation on the hadron structure and the excitation modes has a long history in the particle physics. 
In 1969, Y.~Nambu first pointed out the string picture for hadrons~\cite{N69,N70} 
to explain the Veneziano amplitude~\cite{V68} on hadron reactions and resonances.
Since then, the string picture has been one of the most important 
scenarios for hadrons~\cite{N74,KS75} and has provided many interesting ideas 
in the wide region of the elementary particle physics~\cite{P98}.
For instance, 
the hadronic string creates infinite number of hadron resonances as the vibrational modes, 
and these excitations lead to the Hagedorn ``ultimate'' temperature~\cite{H65,P84}, 
which 
gives an interesting theoretical picture for 
the QCD phase transition.

For real hadrons, the hadronic string has 
a spatial extension like a ``flux-tube''~\cite{N74,CNN79,CI86,SST95}, 
as the result of one-dimensional squeezing of the color-electric flux  
in accordance with color confinement~\cite{N74,SST95}. 
Therefore, the vibrational modes of the hadronic flux-tube should be much more complicated, and 
the analysis of the excitation modes is important to clarify the underlying picture for real hadrons. 

In the language of QCD, such non-quark-origin excitation is called as the ``gluonic excitation''~\cite{TS03,SIT04,JKM03}, and 
is physically interpreted as the excitation of 
the gluon-field configuration in the presence of the quark-antiquark pair or the three quarks 
in a color-singlet state.

In the hadron physics, the gluonic excitation is one of the interesting phenomena 
beyond the quark model, and relates to the hybrid hadrons such as $q\bar qG$ and $qqqG$ in the valence picture. 
For instance, the hybrid meson includes the exotic hadrons with the exotic quantum number such as  
$J^{PC}=0^{--}, 0^{+-}, 1^{-+}, 2^{+-}, \cdots$,
which cannot be constructed in the simple quark model~\cite{DGH92}.
Then, it is important to investigate the gluonic excitation 
with lattice QCD, not only from the theoretical viewpoint
but also from the experimental viewpoint.

In this paper, 
we study the excited-state three-quark (3Q) potential 
and the gluonic excitation in baryons 
using lattice QCD~\cite{TS03},  
to get deeper insight on these excitations beyond the hypothetical models
such as the string and the flux-tube models.
In QCD, the excited-state 3Q potential is defined as
the energy of the excited state of the gluon-field configuration 
in the presence of the static three quarks, and 
the gluonic-excitation energy is expressed as 
the energy difference between the ground-state 3Q potential \cite{TMNS99,TMNS01,TSNM02} 
and the excited-state 3Q potential.

Note that the inter-quark potential is one of the most fundamental quantities 
directly connected to color confinement, 
and plays the important role for hadron properties.
As for the Q-$\rm\bar Q$ system, a lot of lattice QCD studies~\cite{C7980,BS92,B01,LW01} 
have shown that the ground-state Q-$\rm\bar Q$ potential $V_{\rm Q\bar Q}^{\rm g.s.}$
is well described by the Coulomb plus linear potential,  
$
V_{\rm Q\bar Q}^{\rm g.s.}=
-\frac{A_{\rm Q\bar Q}}{r}+\sigma_{\rm Q\bar Q} r+C_{\rm Q\bar Q},
$
with the inter-quark distance $r$. 
Its behavior at short distances can be explained by 
the Coulomb interaction from the one-gluon-exchange (OGE) process, 
and the linear confinement term  
seems to indicate the flux-tube picture~\cite{N74,CNN79}, where 
the quark and the antiquark are linked by 
one dimensional flux-tube with the string tension of 
$\sigma \simeq 0.89$ GeV/fm.

In QCD, the three-body force among three quarks is also 
a ``primary'' force reflecting the SU(3) gauge symmetry, 
while the three-body force is regarded as a residual interaction in most fields in physics.
In fact, the 3Q potential is directly responsible 
for the structure and properties of baryons, 
similar to the relevant role of the Q-$\bar{\rm Q}$ potential for meson properties, 
and both the Q-$\bar{\rm Q}$ potential and 
the 3Q potential are equally important fundamental quantities in QCD.
Furthermore, the 3Q potential is the key quantity to clarify the quark confinement mechanism in baryons.
However, in contrast with the Q-$\rm\bar Q$ potential, 
there were only a few  preliminary lattice-QCD works \cite{SW8486,KEFLM87,TES88} done in 1980's  
for the ground-state 3Q potential before our first study in 1999~\cite{TMNS99}.

\begin{figure}[ht]
\begin{center}
\includegraphics[scale=0.55]{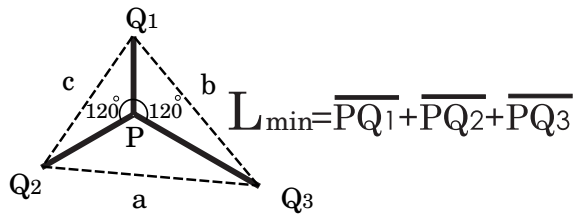}
\end{center}
\caption{The Y-type flux-tube configuration of the 3Q system with the minimal value of the total flux-tube length,  
$L_{\rm min}=\sum_{i=1}^3 \overline{\rm PQ}_i$.
There appears a physical junction linking the three flux tubes at the Fermat point P. 
\label{fig.lmin}}
\end{figure}

Since 1999, we have performed accurate calculations and detailed analyses of 
the ground-state 3Q potential $V_{\rm 3Q}^{\rm g.s.}$ 
in lattice QCD with the smearing method
for more than 300 different patterns of 3Q systems \cite{TSNM02}.
In Refs.~\cite{TMNS99,TMNS01,TSNM02}, 
we have shown that the ground-state 3Q potential $V_{\rm 3Q}^{\rm g.s.}$ 
is well described by the Coulomb plus Y-type linear potential, i.e., the Y-Ansatz, 
\begin{equation}
V_{\rm 3Q}^{\rm g.s.}=-A_{\rm 3Q}\sum_{i<j}\frac1{|{\bf r}_i-{\bf r}_j|}
+\sigma_{\rm 3Q} L_{\rm min}+C_{\rm 3Q}, 
\label{3qpot}
\end{equation}
within 1\%-level deviation. 
Here, $L_{\rm min}$ denotes the minimal value of 
the total length of color-flux-tubes linking the three quarks~\cite{CI86,TMNS99,TMNS01,TSNM02,FRS91,BPV95}, 
which is schematically illustrated in Fig.~\ref{fig.lmin}.
We have found two remarkable features,
the universality of the string tension as 
$\sigma_{\rm 3Q}\simeq \sigma_{\rm Q\bar Q}$ and 
the OGE result as $A_{\rm 3Q}\simeq \frac12 A_{\rm Q\bar Q}$.
(Very recently, we show that the multi-quark potential is well described by the OGE Coulomb 
plus multi-Y type linear potential~\cite{OST04}, which also supports the Y-Ansatz.)

We briefly summarize several other recent studies on the ground-state 3Q potential 
to clarify the current status of the Y-Ansatz. 

After our first study, de Forcrand's group tested only about 20 equilateral 3Q configurations 
in lattice QCD, and supported the $\Delta$-Ansatz \cite{AdFT02}.
However, their data seem to indicate the Y-Ansatz with an overall constant shift, 
and the deviation seems to originate from the data at very short distances,
where the linear potential is negligible compared with the Coulomb contribution.
(Also, their usage of the continuum Coulomb potential may be problematic 
for the lattice data at very short distances.)
Recently, de Forcrand's group seems to change their opinion from the $\Delta$-Ansatz to the Y-Ansatz 
\cite{JdF03}.

One of the theoretical basis of the $\Delta$-Ansatz was Cornwall's conjecture 
based on the vortex vacuum model \cite{C96}. 
Very recently, motivated by our studies, 
Cornwall re-examined his previous work and found an error in the model calculation. 
The corrected answer was the Y-Ansatz instead of the $\Delta$-Ansatz \cite{C04}.

As another analytical work, 
Kuzmenko and Simonov also showed that the Delta-shape is impossible 
from gauge-invariance point of view, 
and the Y-shaped configuration is the only possible for the 3Q 
system \cite{KS03}.

As a clear evidence for the Y-Ansatz, 
Ichie et al. performed the direct measurement on the action density 
in the spatially-fixed 3Q system in lattice QCD 
using the maximally-Abelian projection, 
and observed a clear Y-type flux-tube profile \cite{IBSS03,SIT04}.

In this way, the Y-Ansatz is also supported by various studies of other groups \cite{JdF03,KS03,C04,IBSS03}, and 
therefore the Y-Ansatz for the ground-state 3Q potential is almost settled both in lattice QCD and in analytic framework.

As for the excited-state 3Q potential, however, there is no lattice QCD study besides our previous work~\cite{TS03}.
In this paper, we present the detailed analysis of the excited-state 3Q potential and 
the gluonic excitation 
for about 100 different patterns of the 3Q static systems in SU(3) lattice QCD at the quenched level.

This paper is organized as follows.
In section~\ref{sec2}, we give a necessary formalism 
on the relation between the 3Q Wilson loop and the QCD Hamiltonian. 
We then give the lattice QCD formalism to extract
the excited-state 3Q potential in section~\ref{sec3},
and show the lattice QCD results in section~\ref{sec4}.
In section~\ref{sec5}, 
we investigate the functional form of the gluonic excitation energy
from the lattice QCD data.   
In section~\ref{sec6}, we discuss the physical implication of the obtained lattice results, 
and consider the physical reason of the success of 
the quark model for low-lying hadrons in terms of the gluonic excitation.
Section~\ref{sec7} is devoted to the summary and the conclusion.

\section{3Q Wilson loop and 3Q potentials}\label{sec2}

\subsection{3Q Wilson loop and QCD Hamiltonian}

To begin with, let us consider the physical modes in the static 3Q system.
We denote the $n$th excited state by $|n \rangle$ ($n$=0,1,2,3,..) 
for the physical eigenstates of the QCD Hamiltonian $\hat H$ for the spatially-fixed 3Q system.
For the simple notation, the ground state is expressed as the ``0th excited state'' in this paper.
Since the three quarks are spatially fixed in this case, 
the eigenvalue of $\hat H$ is expressed by the static 3Q potential as 
\begin{eqnarray}
\hat H|n\rangle=E_n|n\rangle=V_n|n\rangle, 
\label{3qstate01}
\end{eqnarray}
where $V_n$ denotes the $n$th excited-state 3Q potential.
We take the normalization condition as $\langle m|n \rangle=\delta_{mn}$.
Note that both $V_n$ and $|n \rangle$ are 
universal physical quantities related to the QCD Hamiltonian $\hat H$.

\begin{figure}[h]
\begin{center}
\includegraphics[scale=0.4]{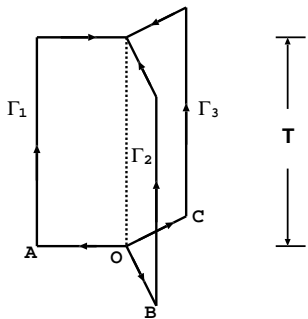}
\end{center}
\caption{\label{fig.3qloop}
The 3Q Wilson loop $W_{\rm 3Q}$.
The gauge-covariant 3Q state $|\Phi\rangle$ is 
generated at $t=0$ and is annihilated at $t=T$. 
The three quarks are spatially fixed in ${\bf R}^3$ for $0 < t < T$.
}
\end{figure}

Similar to the calculation of the Q-${\rm\bar Q}$ potential with the Wilson loop,
the 3Q potential can be calculated with the 3Q Wilson loop $W_{\rm 3Q}$ 
defined as 
\begin{equation}
W_{\rm 3Q} \equiv \frac1{3!}\varepsilon_{abc}\varepsilon_{a'b'c'}
U_1^{aa'} U_2^{bb'} U_3^{cc'} 
\label{3qloop}
\end{equation}
with $U_k \equiv P\exp\{ig\int_{\Gamma_k}dx^\mu A_{\mu}(x)\}$ 
($k=1,2,3$).  
Here, $P$ denotes the path-ordered product along the path denoted by 
${\Gamma_k}$ in Fig.~\ref{fig.3qloop}.

In the 3Q Wilson loop, a gauge-covariant 3Q state $|\Phi\rangle$ is 
generated at $t=0$ and is annihilated at $t=T$ with  
the three quarks being spatially fixed in ${\bf R}^3$ for $0 < t < T$.
In general, the 3Q state $|\Phi\rangle$ in the 3Q Wilson loop 
is not an eigenstate of the QCD Hamiltonian 
$\hat H$ for the spatially-fixed 3Q system, and 
can be expressed with a linear combination of the 3Q physical eigenstates $|n\rangle$ ($n$=0,1,2..) as 
\begin{eqnarray}
|\Phi \rangle =c_0|0\rangle+c_1|1\rangle+c_2|2\rangle+..,
\label{3qstate02}
\end{eqnarray}
where the complex coefficients $c_n$ satisfy  
$\sum_{n=1}^{\infty}|c_n|^2=1$
and express the overlap with each state $|n \rangle$ in the 3Q Wilson loop. 

Since the Euclidean time evolution of the 3Q state 
$|\Phi(t)\rangle$ is expressed with the operator $e^{-\hat Ht}$, 
which corresponds to the transfer matrix in lattice QCD, 
the expectation value of the 3Q Wilson loop is expressed as 
\begin{eqnarray}
&&\langle W_{\rm 3Q}(T)\rangle= \langle\Phi (T)|\Phi (0)\rangle 
=\langle\Phi |e^{-\hat HT}|\Phi \rangle \\ \nonumber
&&=\sum_{m=0}^\infty \sum_{n=0}^\infty \bar c_m c_n
\langle m|e^{-\hat HT}|n \rangle 
= \sum_{n=0}^\infty |c_n|^2 e^{-V_nT}.
\label{3qstate00}
\end{eqnarray}

Then, the low-lying potentials, e.g., 
the ground-state potential $V_{\rm 3Q}^{\rm g.s.}\equiv V_0$ and  
the 1st excited-state potential $V_{\rm 3Q}^{\rm e.s.}\equiv V_1$,  
can be obtained from the large-$T$ behavior of $\langle W_{\rm 3Q}(T)\rangle$ 
where higher excited-state contributions are negligible.
In practical lattice simulations, 
however, it is difficult to calculate $\langle W_{\rm 3Q}(T)\rangle$ accurately for large $T$, 
since its value decreases exponentially with $T$. 

Therefore, for the accurate calculation of low-lying potentials, 
it is desired to reduce higher excited-state components 
in the 3Q state $|\Phi \rangle$ in the 3Q Wilson loop.

\subsection{The smearing method}

For this purpose, we use the smearing method~\cite{APE87} to reduce 
highly excited-state components in the 3Q state $|\Phi \rangle$. 
The smearing is just a method for the state composition in a gauge-covariant manner, 
and never changes physical quantities and gauge configurations, unlike the cooling.

The smearing for link-variables is expressed as the 
iterative replacement of the spatial link-variable $U_i (s)$ 
($i=1,2,3$) by the obscured link-variable 
$\bar U_i (s) \in {\rm SU(3)}$ which maximizes 
\begin{eqnarray}
&&
{\rm Re} \,\,{\rm tr}\Bigg\{ \bar U_i^{\dagger}(s) \Bigg[
\alpha U_i(s)+\sum_{j \ne i} \{
U_j(s)U_i(s+\hat j)U_j^\dagger (s+\hat i)
\nonumber\\
&&\quad\quad\quad+U_j^\dagger (s-\hat j)U_i(s-\hat j)U_j(s+\hat i-\hat j)
\} \Bigg] \Bigg\},
\label{smear}
\end{eqnarray}
leaving the temporal link-variable $U_4(s)$ unchanged.

Note that the smeared 3Q Wilson loop composed by the smeared link-variable 
is expressed as a spatially-extended operator in terms of the original link-variable $U_\mu(s)$, 
and therefore the 3Q state $|\Phi \rangle$ and the coefficients $c_n$ in the smeared 3Q Wilson loop 
are changed according to the iteration number $N_{\rm smr}$ of the smearing. 
In other words, the coefficients $c_n$ in the smeared 3Q Wilson loop 
can be controlled to some extent in the smearing procedure, 
by changing the iteration number $N_{\rm smr}$
and the smearing parameter $\alpha$.

For instance, as the iteration number $N_{\rm smr}$ of the smearing 
increases from $N_{\rm smr}$=0, 
the excited-state components in the $N_{\rm smr}{\rm th}$ 
smeared 3Q Wilson loop gradually decrease, 
and finally the 22th (42th) smeared 3Q Wilson loop is almost ground-state 
saturated as $|\Phi \rangle \simeq c_0^k|0\rangle$
for $\alpha=2.3$ on the lattice with 
$\beta$=5.8 (6.0)~\cite{TSNM02}.
Then, with the properly smeared 3Q Wilson loop $\langle W_{\rm 3Q}(T)\rangle$, 
$V_{\rm 3Q}^{\rm g.s.}$ can be accurately calculated as 
$V_{\rm 3Q}^{\rm g.s.} \simeq -\frac{1}{T}\ln \langle W_{\rm 3Q}(T)\rangle$ 
even at relatively small $T$ \cite{TMNS01,TSNM02}.

The smearing method is also useful to extract the low-lying 3Q potentials, i.e., 
the ground-state potential $V_{\rm 3Q}^{\rm g.s.}$ and  
the 1st excited-state potential $V_{\rm 3Q}^{\rm e.s.}$ \cite{TS03,SIT04}.

\section{Formalism}\label{sec3}

In this section, we present the formalism of the variational method \cite{TS03,LW90,PM90} 
to extract the excited-state potential and the gluonic excitation 
by  diagonalizing the QCD Hamiltonian in the presence of static quarks. 

Suppose that $|\Phi_k \rangle \ (k=0,1,2,3,..)$ 
are arbitrary given independent 3Q states 
for the spatially-fixed 3Q system. In general, each 3Q state  
$|\Phi_k \rangle$ can be expressed with a linear combination 
of the 3Q physical eigenstates $|n\rangle \ (n=0,1,2,..)$ as 
\begin{eqnarray}
|\Phi_k \rangle =c_0^k|0\rangle+c_1^k|1\rangle+c_2^k|2\rangle+..
\label{3qstate03}
\end{eqnarray}
Here, the coefficients $c_n^k$ depend on the selection of the 3Q states 
$|\Phi_k \rangle$, and hence they are not universal quantities.
(Unlike the Q-$\bar{\rm Q}$ system, there is no definite symmetry 
in the 3Q system, so that we do not construct the 3Q states which
carry the specific quantum number.)

The Euclidean time-evolution of the 3Q state 
$|\Phi(t)\rangle$ is expressed with the operator $e^{-\hat Ht}$, 
which corresponds to the transfer matrix in lattice QCD. 
The overlap $\langle \Phi_j(T)|\Phi_k(0)\rangle$ is given by 
the generalized 3Q Wilson loop 
sandwiched by the initial state $\Phi_k$ at $t=0$ 
and the final state $\Phi_j$ at $t=T$, 
and is expressed as the correlation matrix 
in the Euclidean Heisenberg picture: 
\begin{eqnarray}
W^{jk}_T&\equiv& \langle\Phi_j(T)|\Phi_k(0)\rangle 
=\langle \Phi_j|W_{\rm 3Q}(T)|\Phi_k\rangle \nonumber\\
&=&\langle\Phi_j|e^{-\hat HT}|\Phi_k\rangle 
=\sum_{m=0}^\infty \sum_{n=0}^\infty \bar c_m^j c_n^k
\langle m|e^{-\hat HT}|n \rangle \nonumber\\
&=&\sum_{n=0}^\infty \bar c_n^j c_n^k e^{-V_nT}.
\label{3qstate04}
\end{eqnarray}
We define the matrix $C$ and the diagonal matrix $\Lambda_T$ by 
\begin{eqnarray}
C^{nk} =c_n^k, 
\quad \Lambda_T^{mn}=e^{-V_nT}\delta^{mn}, 
\label{3qstate05}
\end{eqnarray}
and rewrite the above relation as 
\begin{eqnarray}
W_T=C^\dagger \Lambda_T C.
\label{3qstate06}
\end{eqnarray}
Note here that $C$ is not a unitary matrix, 
and therefore this relation does not mean 
the simple diagonalization by the unitary transformation, which
connects the two matrices by the similarity transformation.

Since we are interested in the 3Q potential $V_{n} \ (n=0,1,2,..)$ 
in $\Lambda_T$ rather than 
the non-universal matrix $C$, we single out $V_{n}$ 
from the 3Q Wilson loop $W_T$ using the following prescription. 
From Eq.(\ref{3qstate06}), we obtain 
\begin{eqnarray}
W^{-1}_TW_{T+1}&=&\{C^\dagger \Lambda_T C\}^{-1} C^\dagger \Lambda_{T+1} C
\nonumber\\
&=&C^{-1}\Lambda_T^{-1}\Lambda_{T+1}C\nonumber\\
&=&C^{-1}{\rm diag}(e^{-V_0},e^{-V_1},..)C.
\label{3qstate07}
\end{eqnarray}
Now, $W^{-1}_TW_{T+1}$ is expressed by the similarity transformation of the diagonal matrix 
${\rm diag}(e^{-V_0},e^{-V_1},..)$. 
Therefore, $e^{-V_n}$ can be obtained 
as the eigenvalues of the matrix $W_T^{-1}W_{T+1}$, 
i.e., the solutions of the secular equation, 
\begin{eqnarray}
{\rm det}\{W_T^{-1}W_{T+1}-tI\}
&=&{\rm det} \{\Lambda_T^{-1}\Lambda_{T+1}-tI\}\nonumber\\
&=&\prod_{n}
(e^{-V_n}-t)=0, 
\label{3qstate08}
\end{eqnarray}
where $I$ denotes the unit matrix.
The largest eigenvalue corresponds to 
$e^{-V_0}$ and the $n$th largest eigenvalue corresponds to $e^{-V_n}$.

In this way, the 3Q potential $V_n \ (n=0,1,2,..)$ can be obtained from 
the matrix  $W_T^{-1}W_{T+1}$.
In the practical calculation, we prepare $N$ independent sample states 
$|\Phi_k \rangle \ (k=0,1,..,N-1)$. 
If one chooses appropriate states $|\Phi_k \rangle$ 
which does not include highly excited-state components, 
one can truncate the physical states as $|n \rangle \ (n=0,1,2,..,N-1)$. 
Then, $W_T$, $C$ and $\Lambda_T$ are reduced into $N\times N$ matrices, 
and the secular equation Eq.(\ref{3qstate08}) becomes the $N$th order equation. 

\section{Lattice QCD result}
\label{sec4}

In this section, we show the lattice QCD result of the 
1st excited-state 3Q potential 
$V_{\rm 3Q}^{\rm e.s.}$ and the gluonic excitation energy $\Delta E_{\rm 3Q}$ 
as well as the ground-state potential $V_{\rm 3Q}^{\rm g.s.}$ 
for the spatially-fixed static 3Q system. 
The SU(3) lattice QCD calculation is done with the standard plaquette action on 
$16^3 \times 32 $ at $\beta$=5.8 and 6.0 at the quenched level. 
The lattice spacing is found to be $a \simeq$ 0.15 fm at $\beta=5.8$
and $a \simeq$ 0.1 fm at $\beta=6.0$, which 
are set to reproduce the string tension $\sigma$=0.89 GeV/fm 
in the Q-$\bar{\rm Q}$ potential~\cite{SIT04}.
In Table~\ref{tab.gaugeprm},
we summarize the simulation condition and related information 
of the present lattice QCD calculation for the 3Q potentials.

\begin{table}[h]
\caption{
The simulation condition and related information.
For each $\beta$, we list the corresponding lattice spacing $a$, the lattice size, 
the number $N_{\rm 3Q}$ of the different patterns of the 3Q system analyzed, 
the gauge configuration number $N_{\rm conf}$ used for the measurement, 
the number $N_{\rm therm}$ of sweeps for the thermalization,
the number $N_{\rm sep}$ of sweeps for the separation, 
the smearing parameter $\alpha$ and 
the iteration number $N_{\rm smr}$ 
used for the extraction of the 3Q potentials, $V_{\rm 3Q}^{\rm g.s.}$ and $V_{\rm 3Q}^{\rm e.s.}$.
}
\label{tab.gaugeprm}
\begin{tabular}{ c c c c c c c c c } \hline\hline
$\beta$ & $a$[fm] & lattice size & 
$N_{\rm 3Q}$ & $N_{\rm conf}$ &
$N_{\rm therm}$ & $N_{\rm sep}$ &  $\alpha$ & $N_{\rm smr}$
\\ \hline
5.8 & 0.15 & $16^3\times 32$ & 24 & 200 & 10,000 & 500 
& 2.3 & 8,12,16,20 \\
6.0 & 0.10 & $16^3\times 32$ & 73 & 149 & 10,000 & 500 
& 2.3 & 16,24,32,40 \\ 
\hline\hline
\end{tabular}
\end{table}

From now on, we concentrate ourselves on the ground state $|0 \rangle$ and 
the 1st excited state $| 1 \rangle$ in the spatially-fixed 3Q system. 
To extract $V_0$ and $V_1$, we need to prepare at least 
two independent states $|\Phi_k\rangle (k=0,1)$, 
and construct the $2 \times 2$ matrix $W_T^{-1}W_{T+1}$ with them. 
Here, the sample states $|\Phi_k\rangle$  
can be freely chosen, as long as they satisfy the two conditions: 
the linear independence and the smallness of the higher excited-state 
components $|n\rangle$ with $n \ge 2$, 
which leads to $|\Phi_k \rangle \simeq c_0^k|0\rangle+c_1^k|1\rangle$.

As the sample 3Q states $|\Phi_k\rangle$, 
we adopt the properly smeared 3Q states 
since the higher excited-state components are reduced in them~\cite{TSNM02}.
Here,  the smearing parameter is fixed to be $\alpha=2.3$.
After some numerical check on the above two conditions, 
we adopt the 8th, 12th, 16th, 20th smeared 3Q states at $\beta=5.8$
and the 16th, 24th, 32nd, 40th smeared 3Q states at $\beta=6.0$
as the candidates of the sample 3Q states. 
Owing to the intervals of 4 (8) iterations at $\beta$=5.8 (6.0),
these smeared states are clearly independent of each other.
The $N_{\rm smr}$th smeared state  
with $N_{\rm smr}\ge$ 8 (16) at $\beta$=5.8 (6.0) 
has small higher excited-state components.

\begin{figure}[h]
\begin{center}
\includegraphics[scale=0.3]{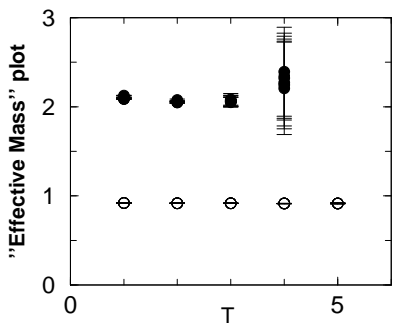}
\caption{\label{fig.efmass}
An example of the effective-mass plot for $V_0(T)$ (open circles) and 
$V_1(T)$ (filled circles) 
obtained from the eigenvalues of $W_T^{-1}W_{T+1}$ at each $T$ 
for the spatially-fixed three quarks 
put on $(1,0,0)$, $(0,1,0)$ and $(0,0,1)$
in the lattice unit at $\beta$=5.8.
For $V_0(T)$ and $V_1(T)$ at each $T$,
we plot 6 data obtained from the 6 pairs of the state combination.
}
\end{center}
\end{figure}

For each possible pairing $(j, k)$ of these 4 states, we calculate the generalized 3Q Wilson loop 
$W_T^{jk} \equiv \langle\Phi_j(T)|\Phi_k(0)\rangle$, 
and evaluate $V_0$ and $V_1$ with Eq.(\ref{3qstate08}). 
We plot in Fig.~\ref{fig.efmass} an example of the ``effective mass'' plot for 
$V_0(T)$ and $V_1(T)$ obtained from Eq.(\ref{3qstate08}) at $\beta=5.8$
as the function of the temporal separation $T$.
For the estimation of the statistical error of the lattice data, 
we adopt the jack-knife error estimate.

For $V_0(T)$ and $V_1(T)$ at each $T$, 
we have 6 ($=\hspace{-0.2cm} \ _4C_2$) data obtained from 6 pairs of the combination among 
the 4 sample states, i.e., the 8th, 12th, 16th, 20th smeared 3Q states. 
As shown in Fig.~\ref{fig.efmass}, these 6 data almost coincide and the 
$T$-dependence of $V_0(T)$ and $V_1(T)$ is 
rather small in a certain region of $T$.
This indicates the smallness of 
the higher excited-state components $|n \rangle$ 
with $n \ge 2$ in the sample states, 
since such contaminations lead a nontrivial $T$-dependence 
in $V_0(T)$ and $V_1(T)$ and make the stability lost.

For the accurate measurement, 
we select the best pairing 
providing the most stable effective-mass plot, 
which physically means the smallest contamination 
of the higher excited states in them. 
(If the effective-mass plot does not show a plateau, 
we exclude the 3Q configuration from the analysis to keep the accuracy.) 
With the selected two states, 
we extract the ground-state potential $V_{\rm 3Q}^{\rm g.s.}$
and the 1st excited-state potential $V_{\rm 3Q}^{\rm e.s.}$
using the $\chi^2$ fit as $V_{\rm 3Q}^{\rm g.s.}=V_0(T)$ 
and $V_{\rm 3Q}^{\rm e.s.}=V_1(T)$ 
in the fit range of $T$, where the plateau is observed.
We perform the above procedure for each 3Q system.

In Tables \ref{tab.onitab00}, \ref{tab.onitab01} and \ref{tab.onitab02}, 
we summarize the lattice QCD results for the ground-state 3Q potential $V_{\rm 3Q}^{\rm g.s.}$ 
and the 1st excited-state potential $V_{\rm 3Q}^{\rm e.s.}$ at the quenched level.
Here, we analyze the 97 different patterns of the spatially-fixed 3Q system in total 
on the $16^3\times 32$ lattice at $\beta$ =5.8 and 6.0.

\begin{table}[h]
\begin{center}
\newcommand{\m}{\hphantom{$-$}}
\newcommand{\cc}[1]{\multicolumn{1}{c}{#1}}
\renewcommand{\tabcolsep}{0.3pc} % enlarge column spacing
\renewcommand{\arraystretch}{1} % enlarge line spacing
\caption{\label{tab.onitab00}
The ground-state 3Q potential $V_{\rm 3Q}^{\rm g.s.}$
and the 1st excited-state 3Q potential $V_{\rm 3Q}^{\rm e.s.}$ 
at $\beta=5.8$ in the lattice unit. 
The label $(l, m, n)$ denotes the 3Q system 
where the three quarks are put on $(la, 0, 0)$, $(0, ma, 0)$ and $(0, 0, na)$ 
in ${\bf R}^3$.
}
\small
\vspace{0.5cm}
\begin{tabular}{c c c c} \hline\hline
$(l, m, n)$ & $V_{\rm 3Q}^{\rm e.s.}$
& $V^{\rm g.s.}_{\rm 3Q}$ & $\Delta E_{\rm 3Q} \equiv V^{\rm e.s.}_{\rm 3Q}-V_{\rm 3Q}^{\rm g.s.}$ \\
\hline
(0,1,1) &  1.9816( 95)  &  0.7711( 3)  &  1.2104( 95)\\
(0,1,2) &  1.9943( 72)  &  0.9682( 4)  &  1.0261( 72)\\
(0,1,3) &  2.0252( 92)  &  1.1134( 7)  &  0.9118( 90)\\
(0,2,2) &  2.0980( 80)  &  1.1377( 6)  &  0.9603( 81)\\
(0,2,3) &  2.1551( 87)  &  1.2686( 9)  &  0.8866( 86)\\
(0,3,3) &  2.2125(114)  &  1.3914(13)  &  0.8211(112)\\
(1,1,1) &  2.0488( 90)  &  0.9176( 4)  &  1.1312( 90)\\
(1,1,2) &  2.0727( 75)  &  1.0686( 5)  &  1.0041( 75)\\
(1,1,3) &  2.1023( 73)  &  1.2004( 7)  &  0.9019( 73)\\
(1,1,4) &  2.1580( 93)  &  1.3201(10)  &  0.8380( 92)\\
(1,2,2) &  2.1405( 72)  &  1.1907( 7)  &  0.9498( 71)\\
(1,2,3) &  2.1899( 71)  &  1.3084( 9)  &  0.8815( 70)\\
(1,2,4) &  2.2516( 79)  &  1.4221(12)  &  0.8296( 78)\\
(1,3,4) &  2.2907( 91)  &  1.5260(15)  &  0.7647( 88)\\
(1,4,4) &  2.3807(138)  &  1.6322(20)  &  0.7485(136)\\
(2,2,2) &  2.1776(111)  &  1.2844(10)  &  0.8932(110)\\
(2,2,3) &  2.2242( 96)  &  1.3882(11)  &  0.8360( 95)\\
(2,2,4) &  2.2799( 98)  &  1.4952(15)  &  0.7847( 99)\\
(2,3,4) &  2.3637(100)  &  1.5853(18)  &  0.7784( 99)\\
(2,4,4) &  2.4108(137)  &  1.6836(23)  &  0.7271(135)\\
(3,3,3) &  2.3408(168)  &  1.5680(19)  &  0.7728(166)\\
(3,3,4) &  2.3958(151)  &  1.6635(22)  &  0.7323(146)\\
(3,4,4) &  2.4645(177)  &  1.7565(30)  &  0.7081(173)\\
(4,4,4) &  2.5245(340)  &  1.8408(42)  &  0.6837(343)\\
\hline\hline
\end{tabular}\\[2pt]
\end{center}
\end{table}

\begin{table}[h]
\begin{center}
\newcommand{\m}{\hphantom{$-$}}
\newcommand{\cc}[1]{\multicolumn{1}{c}{#1}}
\renewcommand{\tabcolsep}{0.3pc} % enlarge column spacing
\renewcommand{\arraystretch}{1} % enlarge line spacing
\caption{\label{tab.onitab01}
The ground-state 3Q potential $V_{\rm 3Q}^{\rm g.s.}$
and the 1st excited-state 3Q potential $V_{\rm 3Q}^{\rm e.s.}$ 
at $\beta=6.0$ in the lattice unit. 
The label $(l, m, n)$ denotes the 3Q system 
where the three quarks are put on $(la, 0, 0)$, $(0, ma, 0)$ and $(0, 0, na)$ 
in ${\bf R}^3$.
}
\small
\vspace{0.5cm}
\begin{tabular}{c c c c} \hline\hline
$(l, m, n)$ & $V_{\rm 3Q}^{\rm e.s.}$
& $V^{\rm g.s.}_{\rm 3Q}$ & $\Delta E_{\rm 3Q} \equiv V^{\rm e.s.}_{\rm 3Q}-V_{\rm 3Q}^{\rm g.s.}$ \\
\hline
(0,1,1 ) &  1.5973(701) &  0.6765( 6) &  0.9209(702)\\
(0,1,2 ) &  1.6502(190) &  0.8233( 6) &  0.8269(190)\\
(0,1,3 ) &  1.6566( 78) &  0.9159( 6) &  0.7407( 77)\\
(0,1,4 ) &  1.6762(101) &  0.9868( 9) &  0.6893(100)\\
(0,1,5 ) &  1.6861( 92) &  1.0491(12) &  0.6370( 91)\\
(0,1,6 ) &  1.7135( 99) &  1.1062(17) &  0.6073( 98)\\
(0,2,2 ) &  1.7380( 75) &  0.9452( 6) &  0.7929( 74)\\
(0,2,3 ) &  1.7559( 79) &  1.0266( 7) &  0.7292( 78)\\
(0,2,5 ) &  1.7858( 88) &  1.1525(13) &  0.6333( 87)\\
(0,2,6 ) &  1.8140( 91) &  1.2084(16) &  0.6055( 90)\\
(0,3,3 ) &  1.7880( 65) &  1.1001( 9) &  0.6879( 64)\\
(0,3,4 ) &  1.8108( 70) &  1.1620(11) &  0.6488( 70)\\
(0,3,5 ) &  1.8325( 78) &  1.2191(15) &  0.6134( 77)\\
(0,3,6 ) &  1.8631( 83) &  1.2721(18) &  0.5910( 83)\\
(0,4,4 ) &  1.8521( 75) &  1.2198(14) &  0.6323( 74)\\
(0,4,5 ) &  1.8843( 77) &  1.2742(17) &  0.6101( 76)\\
(0,4,6 ) &  1.9074( 92) &  1.3281(20) &  0.5793( 91)\\
(0,5,5 ) &  1.9093( 76) &  1.3283(19) &  0.5809( 76)\\
(0,5,6 ) &  1.9450( 84) &  1.3791(24) &  0.5659( 84)\\
(0,6,6 ) &  1.9766(100) &  1.4286(25) &  0.5480( 98)\\
(1,1,1 ) &  1.7035( 66) &  0.7941( 3) &  0.9094( 65)\\
(1,1,2 ) &  1.7236( 77) &  0.8994( 4) &  0.8242( 76)\\
(1,1,3 ) &  1.7196( 74) &  0.9817( 6) &  0.7378( 73)\\
(1,1,4 ) &  1.7348( 97) &  1.0494( 9) &  0.6854( 96)\\
(1,1,5 ) &  1.7422( 84) &  1.1105(12) &  0.6317( 83)\\
(1,2,2 ) &  1.7476( 58) &  0.9810( 5) &  0.7666( 58)\\
(1,2,3 ) &  1.7693( 67) &  1.0521( 7) &  0.7171( 66)\\
(1,2,4 ) &  1.7856( 73) &  1.1151(10) &  0.6705( 71)\\
(1,2,5 ) &  1.8031( 80) &  1.1741(13) &  0.6290( 78)\\
(1,2,6 ) &  1.8273( 87) &  1.2279(16) &  0.5994( 85)\\
(1,3,3 ) &  1.7964( 59) &  1.1161( 9) &  0.6804( 58)\\
(1,3,4 ) &  1.8213( 66) &  1.1745(11) &  0.6467( 64)\\
(1,3,5 ) &  1.8123(202) &  1.2286(22) &  0.5837(202)\\
(1,3,6 ) &  1.8800( 83) &  1.2842(19) &  0.5958( 82)\\
(1,4,4 ) &  1.8483( 66) &  1.2288(14) &  0.6196( 64)\\
(1,4,5 ) &  1.8882( 66) &  1.2829(18) &  0.6053( 64)\\
\hline\hline
\end{tabular}\\[2pt]
\end{center}
\end{table}

\begin{table}[h]
\begin{center}
\newcommand{\m}{\hphantom{$-$}}
\newcommand{\cc}[1]{\multicolumn{1}{c}{#1}}
\renewcommand{\tabcolsep}{0.3pc} % enlarge column spacing
\renewcommand{\arraystretch}{1} % enlarge line spacing
\caption{\label{tab.onitab02}
The ground-state 3Q potential $V_{\rm 3Q}^{\rm g.s.}$
and the 1st excited-state 3Q potential $V_{\rm 3Q}^{\rm e.s.}$ 
at $\beta=6.0$ in the lattice unit. 
The notations are the same as in Table \ref{tab.onitab01}.
}
\small
\vspace{0.5cm}
\begin{tabular}{c c c c} \hline\hline
$(l, m, n)$ & $V_{\rm 3Q}^{\rm e.s.}$
& $V^{\rm g.s.}_{\rm 3Q}$ & $\Delta E_{\rm 3Q} \equiv V^{\rm e.s.}_{\rm 3Q}-V_{\rm 3Q}^{\rm g.s.}$ \\
\hline
(1,5,5 ) &  1.9210( 72) &  1.3343(19) &  0.5867( 71)\\
(1,5,6 ) &  1.9460( 85) &  1.3843(23) &  0.5617( 84)\\
(1,6,6 ) &  1.9855( 88) &  1.4328(26) &  0.5527( 84)\\
(2,2,2 ) &  1.7687( 61) &  1.0392( 6) &  0.7295( 60)\\
(2,2,3 ) &  1.7901( 58) &  1.0993( 8) &  0.6908( 57)\\
(2,2,4 ) &  1.8107( 64) &  1.1575(10) &  0.6532( 63)\\
(2,2,5 ) &  1.8290( 73) &  1.2129(13) &  0.6161( 71)\\
(2,2,6 ) &  1.8631( 76) &  1.2667(17) &  0.5964( 74)\\
(2,3,3 ) &  1.8109( 56) &  1.1509( 9) &  0.6600( 54)\\
(2,3,4 ) &  1.8354( 65) &  1.2044(13) &  0.6310( 64)\\
(2,3,5 ) &  1.8632( 74) &  1.2575(16) &  0.6058( 73)\\
(2,3,6 ) &  1.9053( 77) &  1.3096(18) &  0.5957( 74)\\
(2,4,4 ) &  1.8575( 63) &  1.2547(15) &  0.6028( 62)\\
(2,4,5 ) &  1.9045( 66) &  1.3068(18) &  0.5977( 64)\\
(2,4,6 ) &  1.9167(298) &  1.3542(34) &  0.5625(297)\\
(2,5,5 ) &  1.9295(300) &  1.3509(31) &  0.5786(301)\\
(2,5,6 ) &  1.9689( 76) &  1.4037(24) &  0.5653( 74)\\
(2,6,6 ) &  1.9813( 85) &  1.4493(27) &  0.5320( 80)\\
(3,3,3 ) &  1.8434( 55) &  1.1968(12) &  0.6466( 53)\\
(3,3,4 ) &  1.8695( 60) &  1.2467(14) &  0.6228( 59)\\
(3,3,5 ) &  1.8923( 66) &  1.2963(18) &  0.5961( 63)\\
(3,3,6 ) &  1.9371( 69) &  1.3479(20) &  0.5892( 67)\\
(3,4,4 ) &  1.8464(244) &  1.2879(24) &  0.5584(244)\\
(3,4,5 ) &  1.9164( 67) &  1.3380(20) &  0.5784( 65)\\
(3,4,6 ) &  1.9389( 77) &  1.3881(24) &  0.5507( 71)\\
(3,5,5 ) &  1.9032(322) &  1.3794(34) &  0.5238(319)\\
(3,5,6 ) &  1.9691( 78) &  1.4314(25) &  0.5377( 74)\\
(3,6,6 ) &  1.9743(383) &  1.4717(51) &  0.5026(381)\\
(4,4,4 ) &  1.9215( 61) &  1.3322(18) &  0.5893( 58)\\
(4,4,5 ) &  1.9321( 73) &  1.3766(23) &  0.5555( 72)\\
(4,4,6 ) &  1.9848( 72) &  1.4253(24) &  0.5595( 71)\\
(4,5,5 ) &  1.9569(383) &  1.4190(44) &  0.5379(382)\\
(4,5,6 ) &  1.9749(398) &  1.4582(54) &  0.5166(401)\\
(4,6,6 ) &  1.9667(441) &  1.5039(65) &  0.4628(431)\\
(5,5,6 ) &  2.0076(542) &  1.5050(61) &  0.5026(542)\\
(5,6,6 ) &  2.0150(563) &  1.5445(63) &  0.4705(556)\\
(6,6,6 ) &  2.0970(795) &  1.5925(79) &  0.5046(780)\\
\hline\hline
\end{tabular}
\end{center}
\end{table}

The lattice data of $V_{\rm 3Q}^{\rm g.s.}$ in Tables \ref{tab.onitab00}, \ref{tab.onitab01} and \ref{tab.onitab02} 
also indicate the validity of the Y-Ansatz for the ground-state 3Q potential. 
Note that the present data of $V_{\rm 3Q}^{\rm g.s.}$ are considered to include almost no excited-state contributions, 
since they are extracted by diagonalizing the correlation matrix $\langle \Phi_j|e^{-\hat HT}|\Phi_k\rangle$ 
in terms of the physical basis $|n\rangle$.  
As a consequence, the accuracy of the present data is better than
that of the data in Refs.~\cite{TMNS01,TSNM02}.

\begin{figure}[htb]
\begin{center}
\includegraphics[scale=0.35]{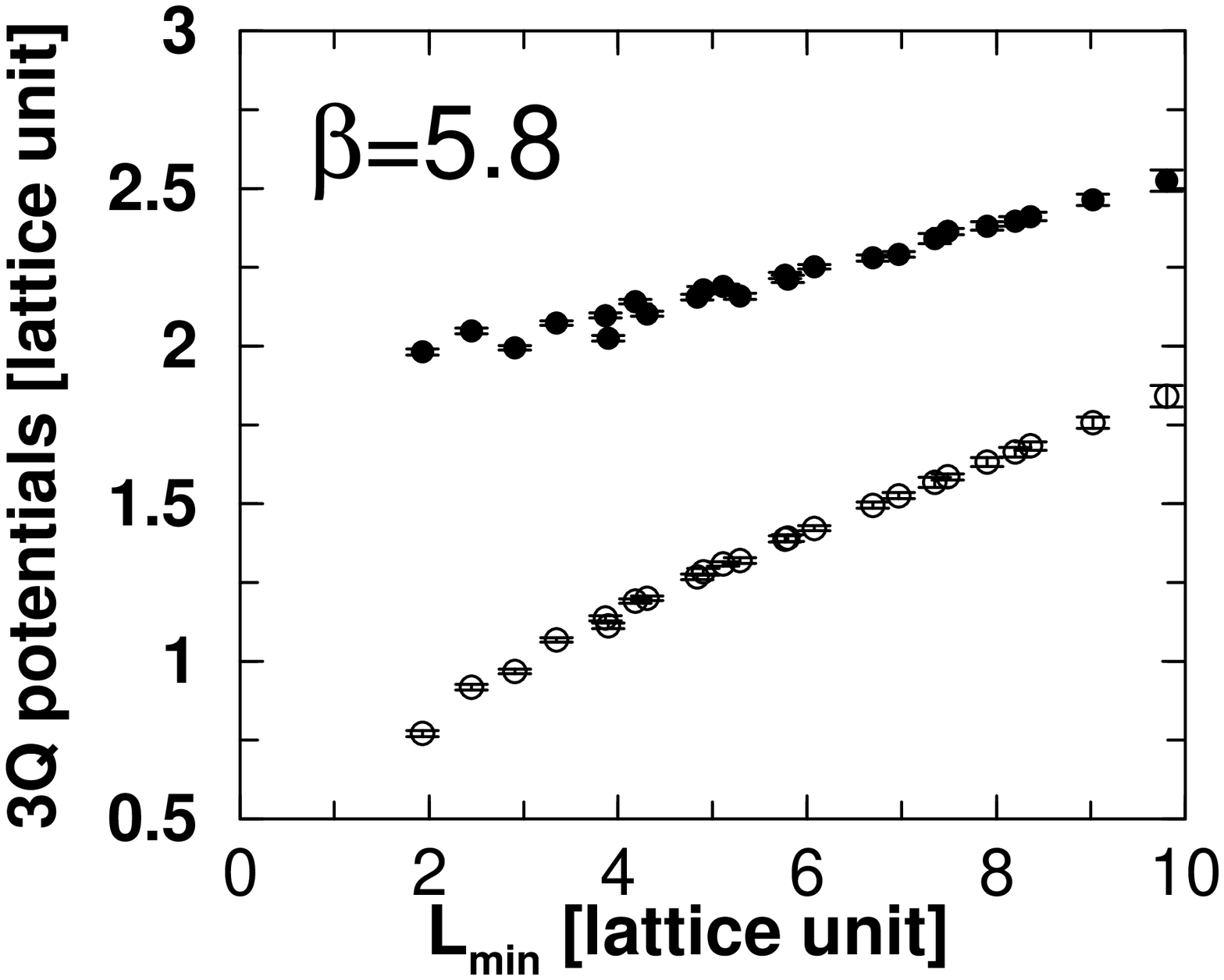}
\includegraphics[scale=0.35]{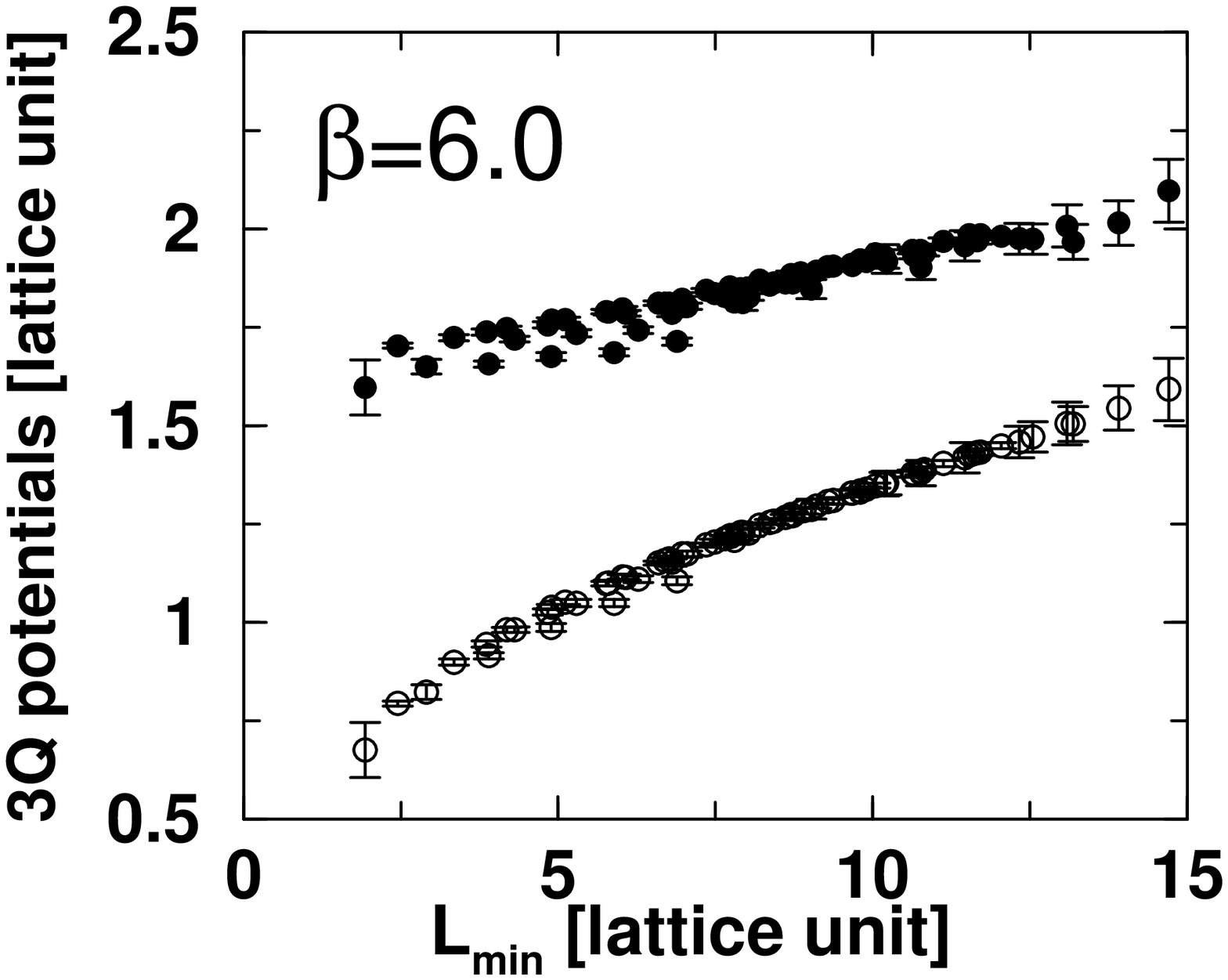}
\end{center}
\caption{\label{fig.3qpots00}The lattice QCD results of 
the ground-state 3Q potential 
$V^{\rm g.s.}_{\rm 3Q}$ (open circles) and the 1st 
excited-state 3Q potential $V^{\rm e.s.}_{\rm 3Q}$ (filled circles)  
plotted against $L_{\rm min}$, 
the minimal total length of flux-tubes linking the three quarks. 
The lattice results at $\beta=5.8$ and $\beta=6.0$ well coincide 
in the physical unit besides an irrelevant overall constant. 
}
\end{figure}

In Fig.~\ref{fig.3qpots00}, 
we plot the ground-state potential $V^{\rm g.s.}_{\rm 3Q}$
and  the 1st excited-state potential $V^{\rm e.s.}_{\rm 3Q}$ 
as the function of the minimal length $L_{\rm min}$ of the Y-type flux-tube
in the lattice unit.
The open symbols are for the ground-state potential 
$V^{\rm g.s.}_{\rm 3Q}$,
and the filled symbols are for the excited-state potential
$V^{\rm e.s.}_{\rm 3Q}$in the 3Q system.
(Note here that we use $L_{\rm min}$ simply for the distinction between 
the different 3Q configurations, and then symbols are not necessarily required
to lie on a single curve in the figures.)
Here, $V^{\rm g.s.}_{\rm 3Q}$ and $V^{\rm e.s.}_{\rm 3Q}$ 
are the lowest and the next-lowest eigenvalues of the QCD Hamiltonian $\hat H$ for the static 3Q system, 
and correspond to the ground-state and the 1st excited-state energies induced by three static quarks in a color-singlet state.
We note that the lattice results at $\beta=5.8$ and $\beta=6.0$ well coincide 
in the physical unit besides an irrelevant overall constant. 
The gluonic excitation energy is expressed as 
$\Delta E_{\rm 3Q} \equiv V^{\rm e.s.}_{\rm 3Q}-V^{\rm g.s.}_{\rm 3Q}$.

It is worth mentioning that the absolute values
of the potentials cannot be determined in lattice QCD without ambiguity.
In fact, the energy of the 3Q system measured with the 3Q Wilson loop
contains an irrelevant constant term $C_{\rm 3Q}$, which 
corresponds to the self-energies of the three static quarks 
under the lattice cutoff $a^{-1}$ and diverges in the 
continuum limit as $a\rightarrow 0$ \cite{TMNS01,TSNM02}.
However, the energy gap between any pair of two states 
does not suffer from the ambiguity and has physical meaning.
In particular, the energy gap $\Delta E_{\rm 3Q} \equiv V_{\rm 3Q}^{\rm e.s.}-V_{\rm 3Q}^{\rm g.s.}$ 
between the ground-state and excited-state potentials 
has definite physical meaning as the lowest gluonic excitation energy, 
and can be determined in lattice QCD without the ambiguity.

\begin{figure}[ht]
\begin{center}
\includegraphics[scale=0.35]{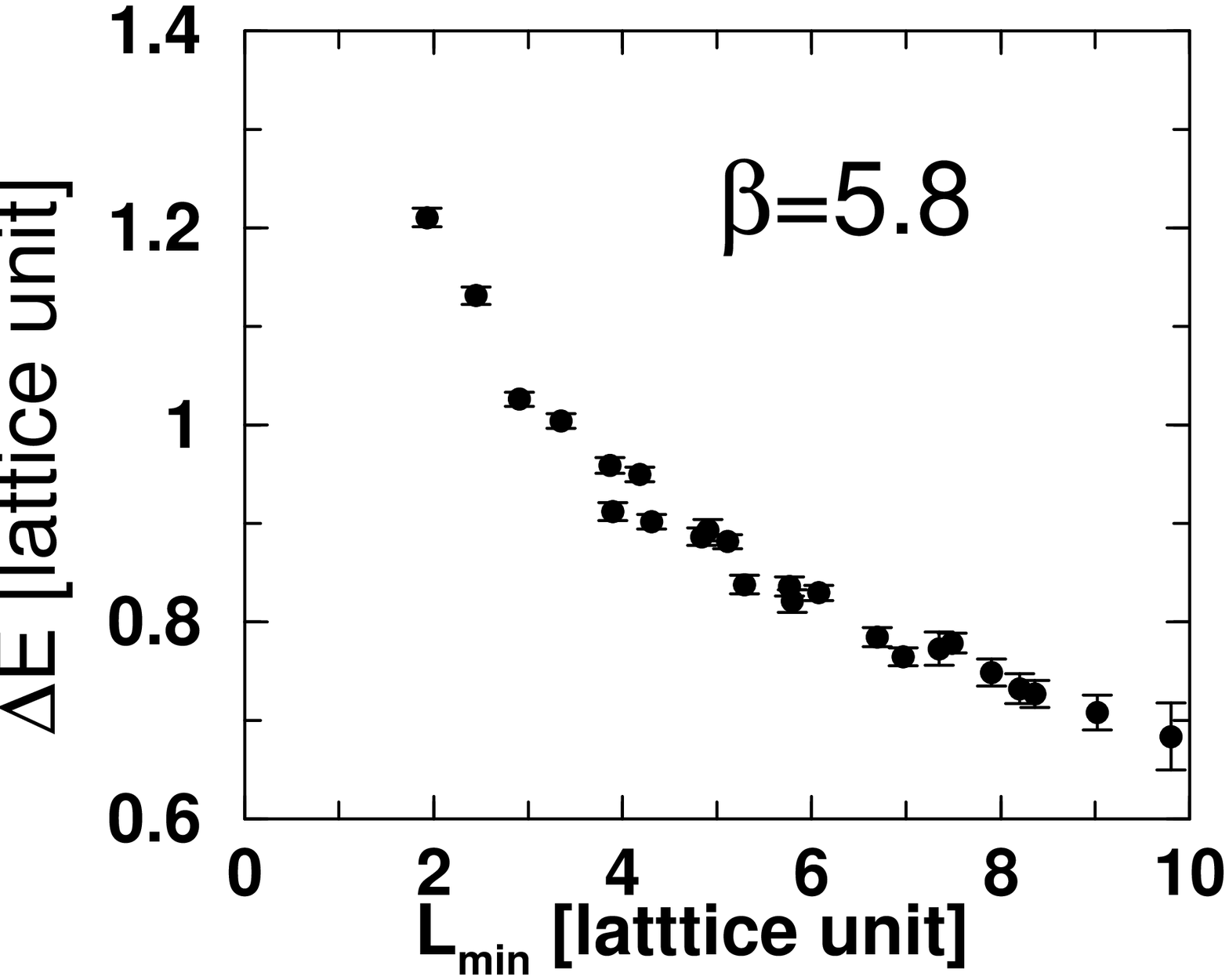}
\includegraphics[scale=0.35]{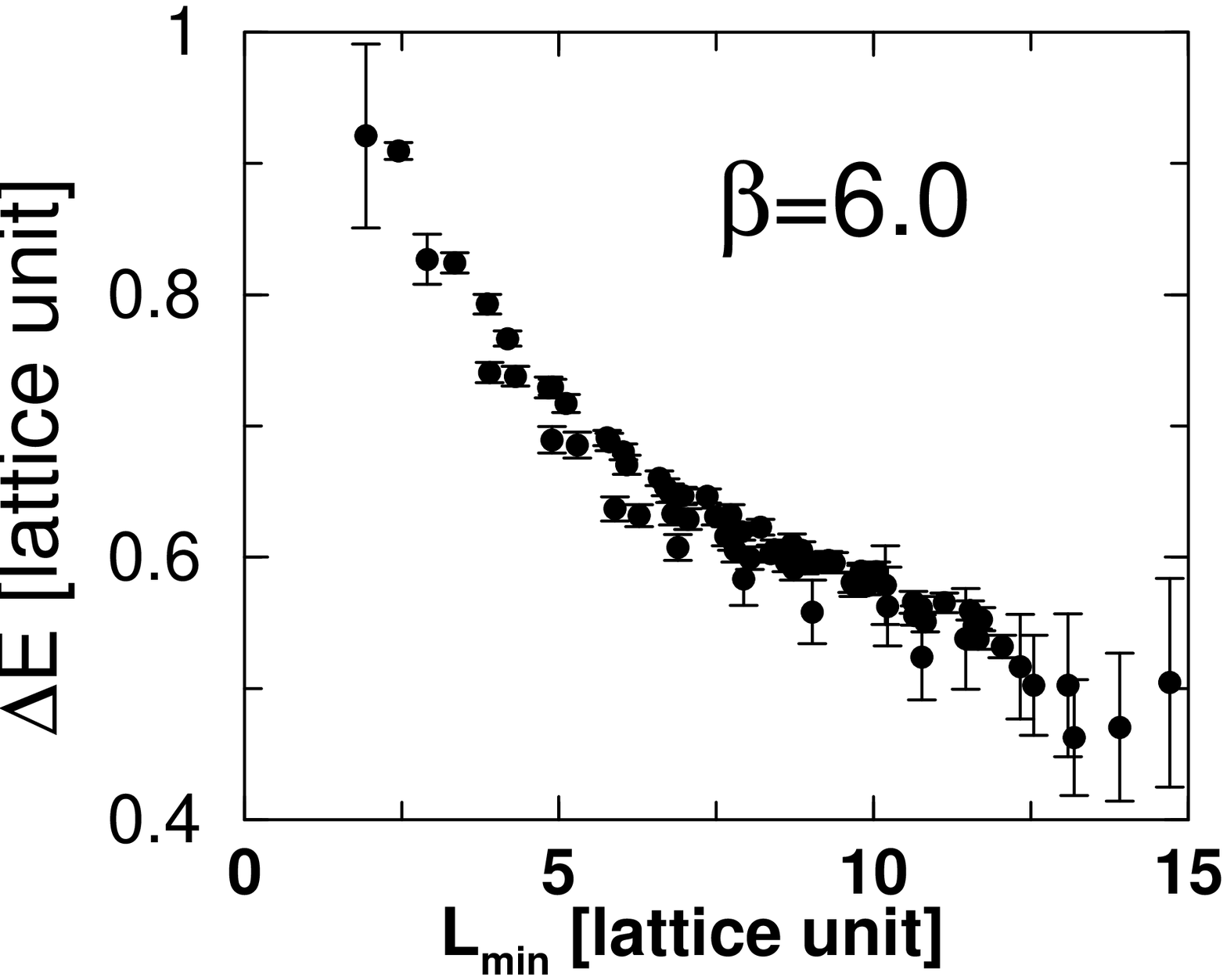}
\end{center}
\caption{\label{fig.3qpots01}
The lattice QCD results of 
the gluonic excitation energy $\Delta E_{\rm 3Q} \equiv V^{\rm g.s.}_{\rm 3Q}-V^{\rm e.s.}_{\rm 3Q}$ 
plotted against $L_{\rm min}$, 
the minimal total length of flux-tubes linking the three quarks. 
The  results at $\beta=5.8$ and $\beta=6.0$ well coincide in the physical unit. 
}
\end{figure}

Figure~\ref{fig.3qpots01} shows 
the gluonic excitation energy $\Delta E_{\rm 3Q} \equiv V_{\rm 3Q}^{\rm e.s.}-V_{\rm 3Q}^{\rm g.s.}$ 
as the function of $L_{\rm min}$. 
As a nontrivial fact, $\Delta E_{\rm 3Q}$ is almost reproduced with a single-valued function of $L_{\rm min}$,
the minimal total length of the flux-tube in the 3Q system.
This implies that the gluonic excitation energy $\Delta E_{\rm 3Q}$
is controlled by the whole size of the 3Q system.
This will be discussed in detail in the next section.

\section{Functional form of gluonic excitation energy}\label{sec5}

In this section, we investigate the functional form of the gluonic excitation energy 
$\Delta E_{\rm 3Q} ({\bf r}_1, {\bf r}_2, {\bf r}_3)$ 
in the static 3Q system in terms of the 3Q location ${\bf r}_i$ ($i$=1,2,3) 
using the lattice QCD data for 
$\Delta E_{\rm 3Q}$ at $\beta$=5.8 and 6.0.

The 3Q static potential $V_{\rm 3Q}$ generally depends on 
the three independent variables which indicate the 3Q triangle, e.g., \{$a,b,c$\} for 
the three sides of the 3Q triangle, 
while the Q-$\rm\bar Q$ potential $V_{\rm Q\bar Q}$ 
depends only on the relative distance $r$. 
Therefore, the search for the functional form of 
$V_{\rm 3Q}^{\rm e.s.}$ or $\Delta E_{\rm 3Q}$ 
is much more difficult than in the Q-$\rm\bar Q$ case.

Furthermore, unlike the ground-state 3Q potential $V_{\rm 3Q}^{\rm g.s.}$, 
there are no clear theoretical candidates for the functional form 
of the excited-state 3Q potential $V_{\rm 3Q}^{\rm e.s.}$ 
or the gluonic excitation energy $\Delta E_{\rm 3Q}$. 
Hence, it is rather difficult to specify their functional forms, 
and we are obliged to perform ``trial and error''.
Here, we consider various possible functional forms, 
and perform the $\chi^2$-fit for the lattice QCD data for each form.

Since the Coulomb part originated from the OGE process is expected to be equal in 
$V_{\rm 3Q}^{\rm e.s.}$ and $V_{\rm 3Q}^{\rm g.s.}$, 
the Coulomb contribution is considered to be cancelled in the combination of  
$\Delta E_{\rm 3Q}\equiv V_{\rm 3Q}^{\rm e.s.}-V_{\rm 3Q}^{\rm g.s.}$.
Then, the functional form of $\Delta E_{\rm 3Q}$ is expected to be simpler than
the excited-state 3Q potential $V_{\rm 3Q}^{\rm e.s.}$.
Therefore, we investigate the gluonic excitation energy $\Delta E_{\rm 3Q}$
in detail instead of $V_{\rm 3Q}^{\rm g.s.}$.

\subsection{Comparison with the Q-$\bf\bar Q$ system}

To begin with, we attempt to express the gluonic excitation energy 
$\Delta E_{\rm 3Q}$ of the 3Q system 
in terms of $\Delta E_{\rm Q \bar Q}(r)$ of the Q-${\rm \bar Q}$ system,
with considering the physical structure of the 3Q system. 
Here, we refer to Ref.~\cite{JKM03} on the lattice QCD data of 
$\Delta E_{\rm Q \bar Q}(r)$ of the Q-${\rm \bar Q}$ system. 

Let us consider the 3Q system as shown in Fig.~\ref{fig.lmin} 
with the quark location ${\rm Q}_i$ ($i$=1,2,3) and the Fermat point P.
We denote the lengths of the three sides by $a, b$ and $c$. 

If the Y-junction exhibits the fixed-edge nature,
the gluonic excitation of the 3Q system 
is expected to resemble that of the Q-$\rm\bar Q$ system, 
since the three static quarks also play the role of the fixed edges.
If it is the case, the lowest gluonic excitation energy $\Delta E_{\rm 3Q}$
in the 3Q system would be expressed as 
\begin{equation}
\Delta E_{\rm 3Q}\simeq \Delta E_{\rm Q\bar Q}
(r={\rm max} (\overline{\rm PQ}_1, \overline{\rm PQ}_2, \overline{\rm PQ}_3)).
\end{equation}

If the excitation mode can be expressed as a vibrational mode on one side of the 3Q triangle, 
the lowest gluonic excitation is expressed by the lowest vibrational mode 
of the Q-$\rm\bar Q$ flux as 
\begin{equation}
\Delta E_{\rm 3Q}\simeq \Delta E_{\rm Q\bar Q}
(r={\rm max} (a,b,c)).
\end{equation}

However, these fits cannot reproduce the lattice QCD data
of $\Delta E_{\rm 3Q}$ at all.
Then, the 3Q gluonic excitation is considered as the bulk excitation of the whole 3Q system,
rather than the excitation of its partial system.

We have also checked many trial forms with $\Delta E_{\rm Q \bar Q}(r)$ such as  
\begin{eqnarray}
\Delta E_{\rm 3Q}&\simeq& \Delta E_{\rm Q\bar Q}(r=L_{\rm min}), \\
\Delta E_{\rm 3Q}&\simeq& \Delta E_{\rm Q\bar Q}(r=a+b+c).
\end{eqnarray}
However, all of them fail to reproduce the lattice QCD result of 
$\Delta E_{\rm 3Q}$.

\subsection{The inverse Mercedes Ansatz}

Next, as a trial, we attempt to plot the gluonic excitation $\Delta E_{\rm 3Q}$ in the 3Q system against 
the minimal total length of the flux-tubes linking the three quarks, 
$L_{\rm min}= \overline{\rm PQ}_1+\overline{\rm PQ}_2+\overline{\rm PQ}_3$, 
as shown in Fig.~\ref{fig.lmin}.

As a remarkable fact, $\Delta E_{\rm 3Q}$
seems to be relatively well expressed as a 
single-valued function of $L_{\rm min}$.
Indeed, although there is some visible deviation, 
the lattice QCD data for $\Delta E_{\rm 3Q}$
nearly collapse to a single curve in Fig.~\ref{fig.3qpots01}.
This is rather nontrivial because $\Delta E_{\rm 3Q}$ 
depends not only on $L_{\rm min}$ 
but also on three independent variables.

\begin{figure}
\includegraphics[scale=0.3]{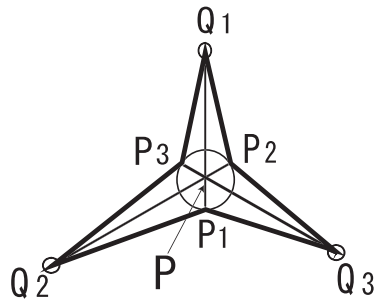}
\caption{
The ``Mercedes form" for the 3Q system, 
where quarks are put on ${\rm Q}_i$ ($i$=1,2,3). 
The modified Y-type flux-tube length $L_{\rm \bar Y}$ 
is expressed as $L_{\rm \bar Y} = \frac12\sum_{i\neq j}\overline{{\rm P}_i{\rm Q}_j}$.
P denotes the Fermat point of $\triangle {\rm Q_1Q_2Q_3}$, and 
${\rm P}_i$ is taken on the extended line of ${\rm Q}_i{\rm P}$ 
so as to satisfy $\overline{\rm PP}_i=\xi$.
The ``inverse Mercedes Ansatz" is defined as $\Delta E_{\rm 3Q}=K/L_{\rm \bar Y}+G$.
}
\label{fig.mody}
\end{figure}

For further investigation, 
we consider the ``Mercedes form" for the 3Q system as shown in Fig.~\ref{fig.mody}. 
We define $x_i \equiv \overline{\rm PQ}_i$
as the distance between the Fermat point P and each quark location ${\rm Q}_i$.

After some trials, 
we finally find that the ``inverse Mercedes Ansatz'' defined by the following functional form
well reproduces the lattice QCD results 
for the gluonic excitation energy $\Delta E_{\rm 3Q}$
in the 3Q system:
\begin{eqnarray}
\Delta E_{\rm 3Q} &=&\frac{K}{L_{\rm\bar Y}}+G, \label{deltae} \\
L_{\rm\bar Y}&\equiv& {\sum_{i=1}^{3}\sqrt{x_i^2-\xi x_i+\xi^2}} \quad
(x_i \equiv \overline{\rm PQ}_i)
\label{lybar}
\end{eqnarray}
with three parameters, $K$, $G$ and $\xi$. 

Here, we refer to $L_{\rm \bar Y}$ as the ``modified Y-type flux-tube length'', or the ``modified Y-length'' simply.
For $\xi =0$, $L_{\rm \bar Y}$ coincides with the Y-type flux-tube length $L_{\rm min}=
\sum_{i=1}^{3}\overline{\rm PQ}_i=\sum_{i=1}^{3}x_i$.
Note that $2L_{\rm\bar Y}$ expresses the total perimeter of the ``Mercedes form" shown in Fig.~\ref{fig.mody}, i.e.,
\begin{equation}
L_{\rm \bar Y}=
\frac12
\sum_{i\neq j}\overline{{\rm P}_i{\rm Q}_j},
\end{equation}
since $\overline{\rm P_2Q_1}^2=
x_1^2+\xi^2-2\xi x_1\cos\frac{\pi}{3}
=x_1^2-\xi x_1+\xi^2$ etc.

\begin{figure}[htb]
\begin{center}
\includegraphics[scale=0.35]{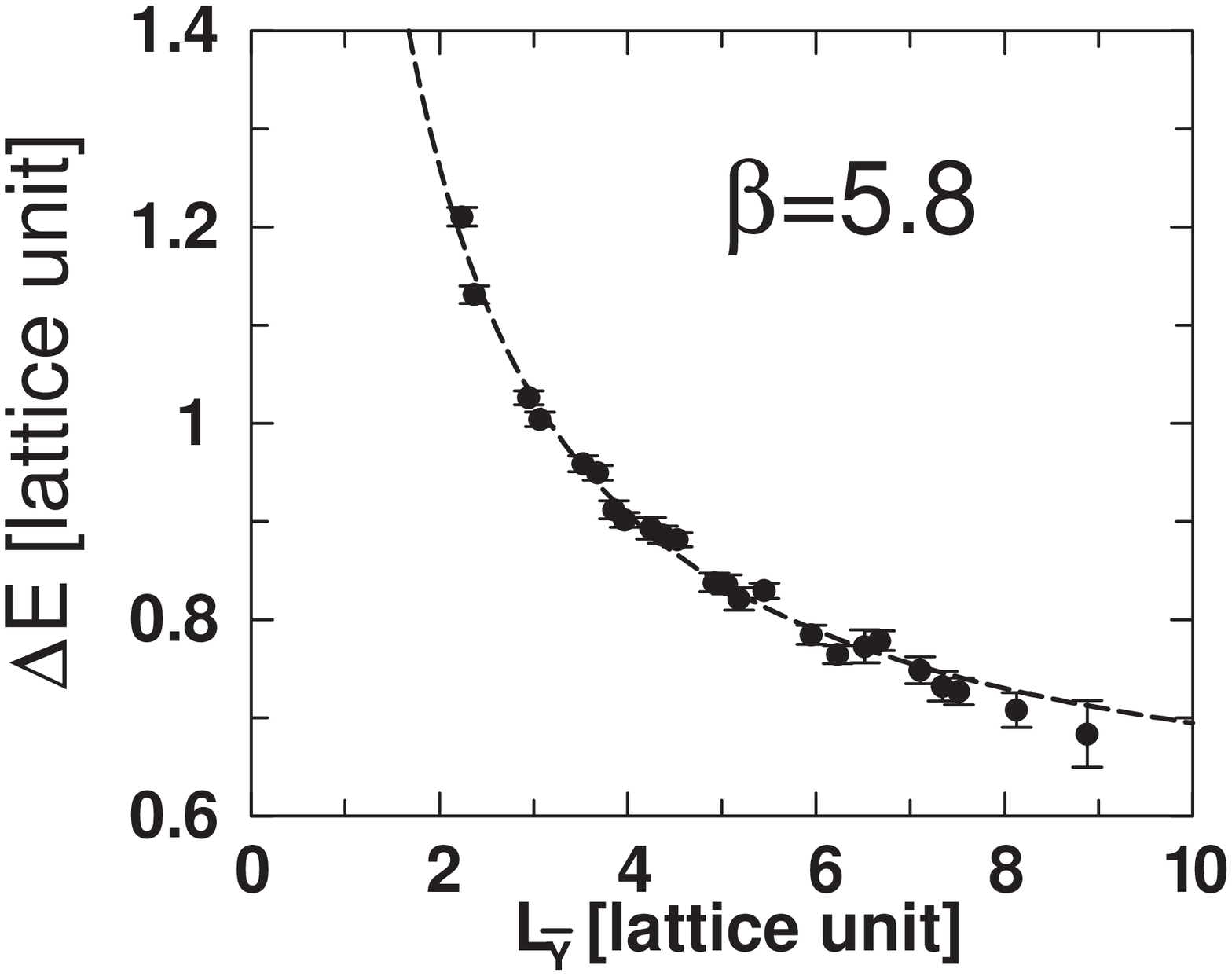}
\includegraphics[scale=0.35]{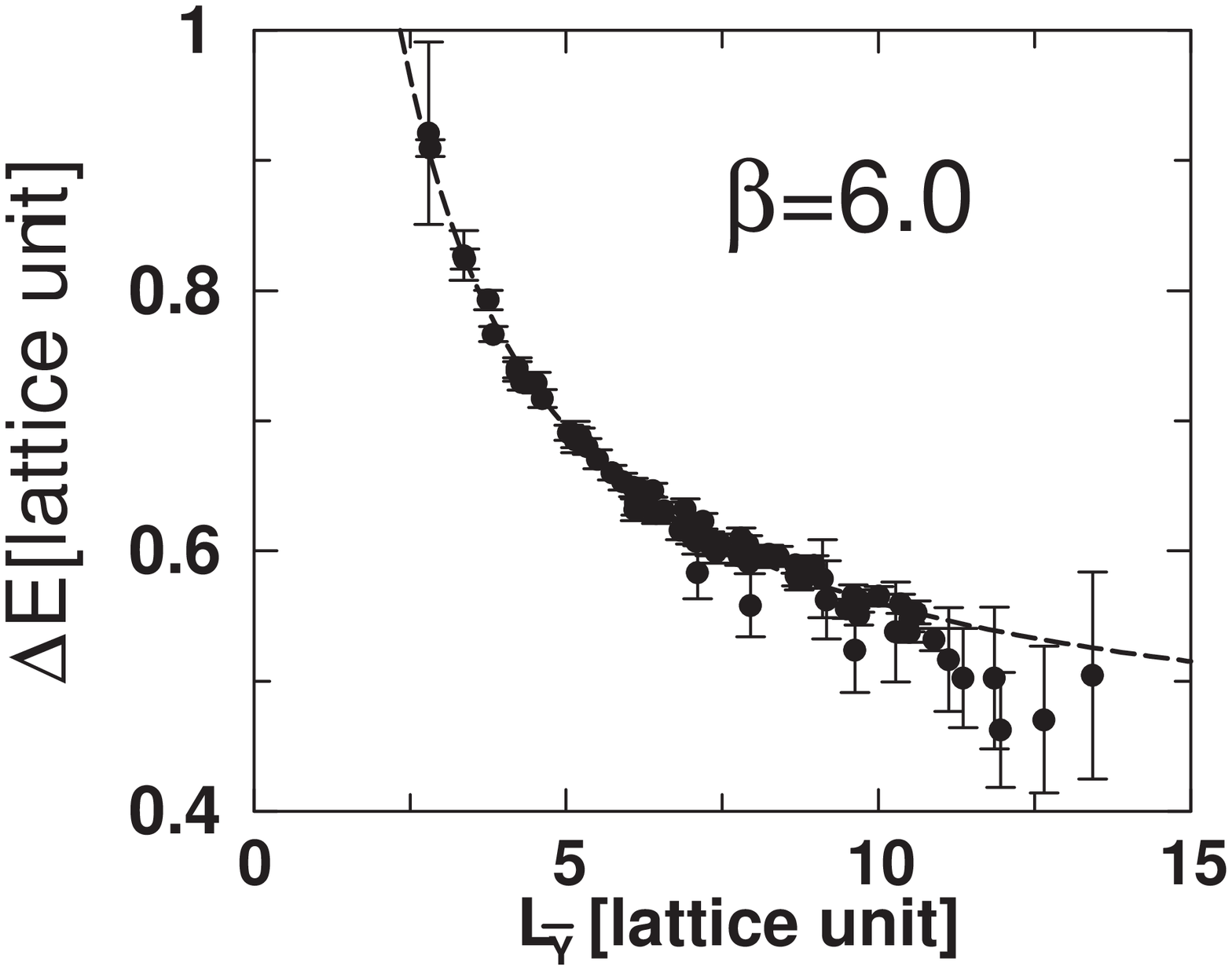}
\end{center}
\caption{\label{fig.3qpots02}
The lattice QCD results of the gluonic excitation energy $\Delta E_{\rm 3Q} \equiv V^{\rm g.s.}_{\rm 3Q}-V^{\rm e.s.}_{\rm 3Q}$
plotted against the modified Y-length $L_{\overline{\rm Y}}$ defined in Eq.(\ref{lybar}).
The dashed curve denotes the inverse Mercedes Ansatz with the best-fit parameters listed in Table \ref{tab.fitprm}.  
}\label{IMa}
\end{figure}

\begin{figure}[htb]
\begin{center}
\includegraphics[scale=0.35]{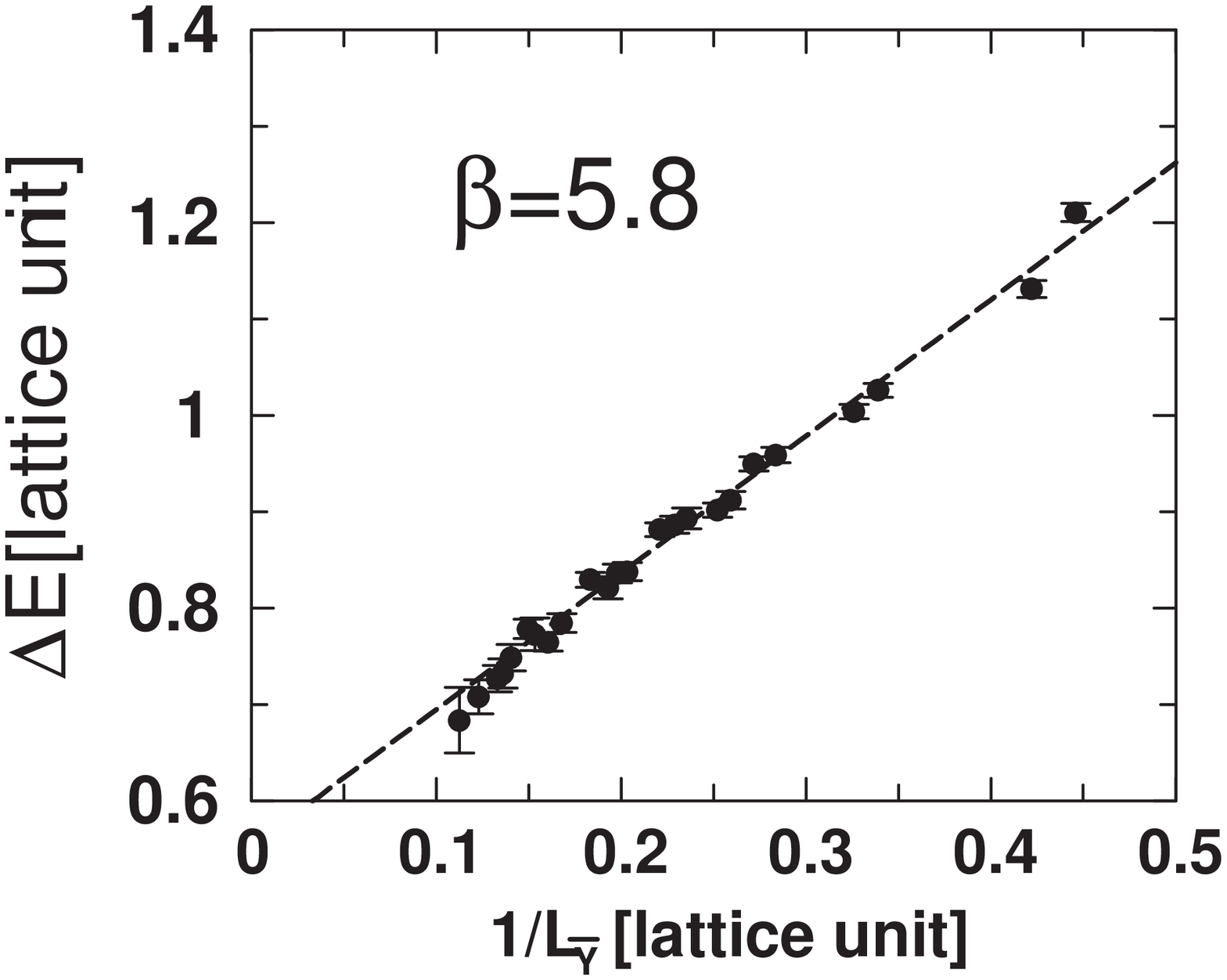}
\includegraphics[scale=0.35]{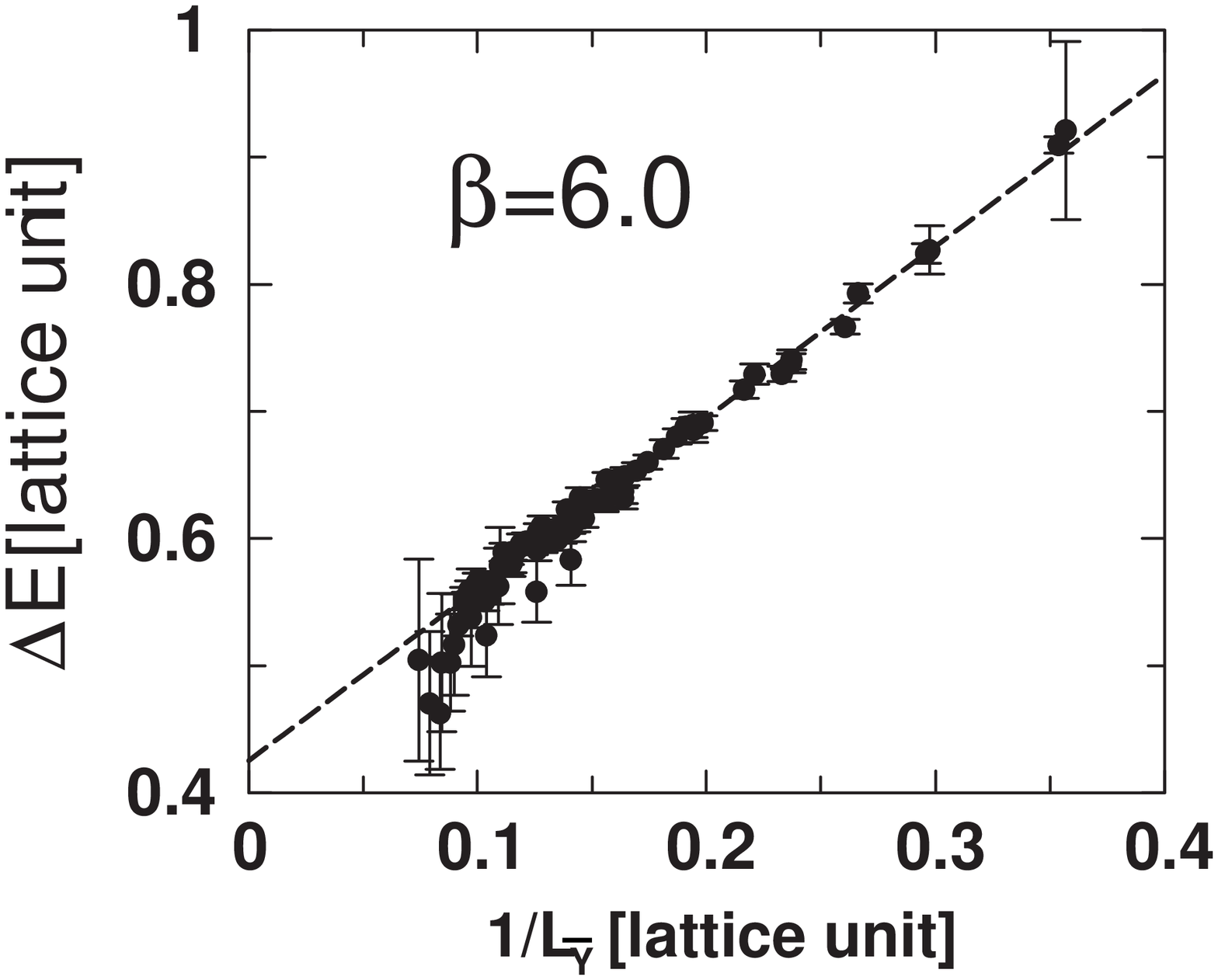}
\end{center}
\caption{\label{fig.3qpots03}
The lattice QCD results of the gluonic excitation energy $\Delta E_{\rm 3Q} \equiv V^{\rm g.s.}_{\rm 3Q}-V^{\rm e.s.}_{\rm 3Q}$
plotted against $1/L_{\overline{\rm Y}}$.
The dashed line denotes the inverse Mercedes Ansatz with the best-fit parameters listed in Table \ref{tab.fitprm}.  
}\label{IMb}
\end{figure}

In Fig.~\ref{IMa}, we plot 
the lattice QCD results of the gluonic excitation energy $\Delta E_{\rm 3Q} \equiv V^{\rm g.s.}_{\rm 3Q}-V^{\rm e.s.}_{\rm 3Q}$
against the modified Y-length $L_{\overline{\rm Y}}$ defined in Eq.(\ref{lybar}).
As a remarkable fact, $\Delta E_{\rm 3Q}$ can be plotted as a single-valued function of $L_{\rm \bar Y}$.
In Fig.~\ref{IMb}, we also plot $\Delta E_{\rm 3Q} \equiv V^{\rm g.s.}_{\rm 3Q}-V^{\rm e.s.}_{\rm 3Q}$
against $1/L_{\overline{\rm Y}}$, since the inverse Mercedes Ansatz 
corresponds to a linear arising behavior in this plot.

We summarize in Table~\ref{tab.fitprm}
the fit analysis with the inverse Mercedes Ansatz for 
the lattice QCD data of the gluonic excitation energy 
$\Delta E_{\rm 3Q}$ at each $\beta$ 
together with the best-fit parameter set, ($K$, $G$, $\xi$).
(In Appendix, we summarize various fit analyses for $\Delta E_{\rm 3Q}$ 
with several trial fit functions of \{$a,b,c$\} or \{$x_1,x_2,x_3$\}.)

\begin{table}[htb]
\caption{\label{tab.fitprm}
The fit analysis with the ``inverse Mercedes Ansatz" for 
the lattice QCD data of the gluonic excitation energy 
$\Delta E_{\rm 3Q}$  at $\beta$=5.8 and 6.0.
The best-fit parameter set ($K$, $G$, $\xi$) is listed in the lattice unit.
One finds $G(\beta=5.8) \simeq$ 0.77 GeV, $G(\beta=6.0) \simeq$ 0.85 GeV, 
$\xi(\beta=5.8) \simeq$ 0.116 fm and $\xi(\beta=6.0) \simeq$ 0.103 fm in the physical unit.
}
\begin{tabular}{c c c c c}
\hline\hline
$\beta$ & $K$ & $G$ & $\xi$ & $\chi^2/N_{DF}$ \\ \hline
$5.8$ & 1.4329(340) &  0.5500(105) &  0.77 & 41.2/(24-3)=1.96 \\
$6.0$ & 1.3486(277) &  0.4252( 68) &  1.03 & 76.8/(73-3)=1.10 \\ \hline
\end{tabular}
\end{table}

In Figs.~\ref{IMa} and \ref{IMb}, we have added by the dashed line the best-fit result of 
the inverse Mercedes Ansatz with the parameters listed in Table \ref{tab.fitprm}.  
The inverse Mercedes Ansatz seems to reproduce fairly well 
the lattice QCD data for the 3Q gluonic excitation energy
$\Delta E_{\rm 3Q}$
at least for the short and the intermediate distances
as $L_{\rm min}\leq$ 2~fm. 

Note again that $\Delta E_{\rm 3Q}$ is generally a function
of the three variables \{$x_1$, $x_2$, $x_3$\}.
However, the inverse Mercedes Ansatz (\ref{deltae})  
depends only on $L_{\rm \bar Y}$, 
which is a single symmetric combination of \{$x_1$,$x_2$,$x_3$\}.
We stress that such a simple dependence
is rather nontrivial.

Now, we pay attention to the parameters $K$, $G$ and $\xi$.
As the physical meaning of the parameters $\xi$ and $G$ in the inverse Mercedes Ansatz, 
$\xi$ plays the role of an ``ultraviolet cutoff'' parameter when all of $x_i$ are small as $x_i < \xi$, 
and the parameter $G$ seems to give an infrared value of the gluonic excitation energy.
We find a relatively good scaling behavior for the parameter set $(K,G,\xi)$ in Table~\ref{tab.fitprm}:
$K(\beta=5.8) \simeq 1.43$, $K(\beta=6.0) \simeq 1.35$, 
$G(\beta=5.8) \simeq$ 0.77 GeV, $G(\beta=6.0) \simeq$ 0.85 GeV, 
$\xi(\beta=5.8) \simeq$ 0.116 fm and $\xi(\beta=6.0) \simeq$ 0.103 fm in the physical unit.

As a caution, one has to be careful for the argument on the infrared behavior
of $\Delta E_{\rm 3Q}$.
For, the character of the gluonic excitation mode 
may be changed into the stringy behavior in the infrared region,
as is actually shown in the Q-$\rm\bar Q$ 
gluonic excitation mode for $r\geq 2$ fm.
For the definite conclusion of the infrared behavior
of $\Delta E_{\rm 3Q}$,
we need to perform the lattice QCD calculation for $\Delta E_{\rm 3Q}$
with much larger lattice volume.

\section{Discussion}\label{sec6}

\subsection{Physical implication of the lattice QCD results}

We consider the physical meaning of the present lattice QCD result, 
although the precise physical interpretation of the inverse Mercedes Ansatz (\ref{deltae})
would be rather difficult and is an open problem.
Since the inverse Mercedes Ansatz is described with the modified Y-type flux-tube length $L_{\rm \bar Y}$,  
the gluonic excitation would be regarded as a global excitation of the whole Y-type flux-tube system,
instead of the partial excitation of each flux-tube as
${\rm PQ}_1$, ${\rm PQ}_2$ or ${\rm PQ}_3$. 
This would exclude the quasi-fixed edge nature of the Y-type junction, 
as was also indicated in section\ref{sec5}-A.
In fact, the inverse Mercedes Ansatz indicates that  
the gluonic excitation in the 3Q system appears
as a complicated excitation of the whole 3Q system.

As a remarkable fact on the absolute value of the gluonic excitation energy,
the lowest gluonic-excitation energy $\Delta E_{\rm 3Q}$
is found to be about 1 GeV or more in the typical hadronic scale as 0.5 fm$\leq L_{\rm min}\leq$ 1.5 fm. 
In fact, the gluonic excitation energy $\Delta E_{\rm 3Q}$ 
is rather large in comparison with the low-lying excitation energy of the quark origin.
(Also for the Q-${\rm \bar Q}$ system, a large gluonic excitation
energy is reported in recent lattice studies~\cite{JKM03}.)
Therefore, the contribution of gluonic excitations 
is considered to be negligible and the dominant contribution is brought 
by quark dynamics such as the spin-orbit interaction for low-lying hadrons.

On the other hand, the gluonic excitation would be significant and visible 
in the highly-excited baryons with the excitation energy above 1 GeV.
For instance, the lowest hybrid baryon~\cite{CP02}, which is described as $qqqG$ in
the valence picture, is expected to have a large mass of about 2 GeV. 
This lattice QCD result may suggest a large ``constituent gluon mass'' 
of about 1 GeV in the constituent gluon picture.

\subsection{Gluonic excitation and success of quark model}

We consider the connection between QCD and the quark model 
in terms of the gluonic excitation \cite{TS03,SIT04}.
While QCD is described with quarks and gluons, 
the simple quark model, which contains only quarks as explicit degrees of
freedom, successfully describes low-lying hadrons,  
in spite of the absence of gluonic excitation modes and the non-relativistic treatment.
As for the non-relativistic treatment, it is conjectured to be justified by 
a large mass generation of quarks due to dynamical chiral-symmetry breaking (DCSB).
However, the absence of the gluonic excitation modes in low-lying hadron spectra 
has been a puzzle in the hadron physics.

On this point, we find the gluonic-excitation energy to be about 1 GeV or more, 
which is rather large compared with the excitation energies of the quark origin,
and therefore the effect of gluonic excitations is negligible 
and quark degrees of freedom plays the dominant role 
in low-lying hadrons with the excitation energy below 1 GeV.
Thus, the large gluonic-excitation energy of about 1 GeV gives the physical reason for 
the invisible gluonic excitation in low-lying hadron spectra, 
which would play the key role to the success of the quark model 
without gluonic excitation modes~\cite{TS03,SIT04}. 

\begin{figure}[ht]
\centerline{\includegraphics[width=0.475\textwidth]{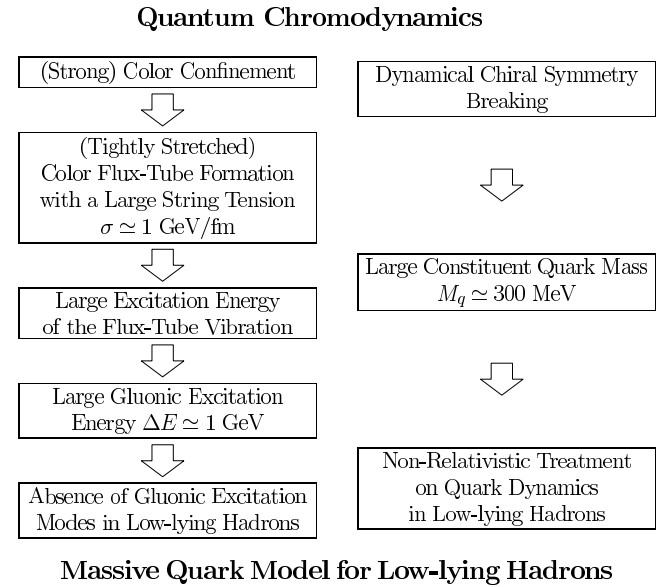}}
\caption{A possible scenario from QCD to the quark model in terms of 
color confinement and DCSB.
DCSB leads to a large constituent quark mass of about 300 MeV, which enables the non-relativistic treatment for quark dynamics approximately. 
Color confinement results in the color flux-tube formation among quarks with a large string tension of $\sigma \simeq$ 1 GeV/fm.
In the flux-tube picture, the gluonic excitation is described as the flux-tube vibration, 
and the flux-tube vibrational energy is expected to be large, 
reflecting the large string tension.
The large gluonic-excitation energy of about 1 GeV leads to 
the absence of the gluonic mode in low-lying hadrons, 
which would play the key role to the success of the quark model without gluonic excitation modes.
}
\label{Quark}
\end{figure}

In Fig.~\ref{Quark}, by way of the flux-tube picture, 
we present a possible scenario from QCD to the massive quark model 
in terms of color confinement and DCSB~\cite{SIT04}.
On one hand, DCSB gives rise of a large constituent quark mass of about 300 MeV, 
which enables the non-relativistic treatment for quark dynamics approximately. 
On the other hand, color confinement results in the color flux-tube formation 
among quarks with a large string tension of $\sigma \simeq$ 1 GeV/fm.
In the flux-tube picture, the gluonic excitation is described as the flux-tube vibration, 
and the flux-tube vibrational energy is expected to be large, reflecting the large string tension 
and the resulting short string.
Due to the large gluonic-excitation energy, which is actually estimated as about 1 GeV in lattice QCD, 
the gluonic excitation seems invisible in low-lying hadrons.
In this way, the quark model becomes successful even without explicit gluonic modes.

\section{Summary and conclusion}\label{sec7}

For about 100 different patterns of spatially-fixed three-quark (3Q) systems, 
we have studied the excited-state 3Q potential and the gluonic excitation 
using SU(3) lattice QCD with $16^3\times 32$ at $\beta$=5.8 and 6.0 at the quenched level. 
We have extracted the excited-state potential 
$V_{\rm 3Q}^{\rm e.s.}$ together with the ground-state potential $V_{\rm 3Q}^{\rm g.s.}$ 
by diagonalizing the QCD Hamiltonian in the presence of three quarks, 
for 24 patterns of 3Q systems at $\beta=5.8$ and for 73 patters at $\beta=6.0$. 

We have found that the lowest gluonic excitation energy $\Delta E_{\rm 3Q} \equiv V_{\rm 3Q}^{\rm e.s.}-V_{\rm 3Q}^{\rm g.s.}$ 
takes a large value of about 1 GeV at the typical hadronic
scale as 0.5 fm $ \le L_{\rm min}\le$ 1.5 fm.
Therefore, we have conjectured that the ``hybrid baryon'' $qqqG$, which corresponds to  
the gluonic excitation mode, appears as the highly-excited baryon 
with the excitation energy of above 1 GeV.

Next, we have investigated the functional form of the gluonic excitation energy $\Delta E_{\rm 3Q}$ 
in terms of the 3Q location. 
After some trials with various functions, 
we have found that the lattice data of $\Delta E_{\rm 3Q}$ are relatively well reproduced by the ``inverse Mercedes Ansatz'', 
$\Delta E_{\rm 3Q}=K/L_{\rm \bar Y}+G$ with
the ``modified Y-type flux-tube length'' $L_{\rm \bar Y}$.
This nontrivial behavior of $\Delta E_{\rm 3Q}$  seems to indicate that 
the gluonic-excitation mode is realized as a complicated bulk excitation 
of the whole Y-type flux-tube system, instead of the partial excitation of each flux-tube. 

Finally, we have considered the physical consequence of the large gluonic-excitation energy, and 
have presented a possible scenario to give a physical reason of  
the success of the quark model for low-lying hadrons even without explicit gluonic modes.

These lattice QCD data of the excited-state potential 
would be useful for the QCD-based construction of the refined quark model,
which can deal with the gluonic excitation modes in hadrons.
Our results would be also helpful for the comprehension 
of the nature of the ``QCD string''.

\begin{acknowledgments}
H.S. thank Prof. J.M. Cornwall for useful discussions.
H.S. was supported in part by a Grant for Scientific Research 
(No.16540236) from the Ministry of Education, 
Culture, Science and Technology, Japan. 
T.T.T. was supported by the Japan Society for the Promotion of Science (JSPS) for Young Scientists.
The lattice QCD Monte Carlo calculations have been performed on NEC-SX5 at Osaka University 
and on HITACHI-SR8000 at KEK.
\end{acknowledgments}

\appendix

\section{Fit analyses for the gluonic excitation energy}

In Appendix, we summarize various fit analyses
for the gluonic excitation energy in the spatially-fixed 3Q system, 
$\Delta E_{\rm 3Q}\equiv V_{\rm 3Q}^{\rm e.s.}-V_{\rm 3Q}^{\rm g.s.}$, 
in terms of the 3Q spatial configuration.
We examine various trial fit functions of \{$a,b,c$\} or \{$x_1,x_2,x_3$\} 
including two or three free parameters, $a_i$: 
Form 1-10 expressed by Eqs.(A1)-(A10).
(\{$a,b,c$\} and \{$x_1,x_2,x_3$\} are defined in section\ref{sec5}.)

{Form~1} is the ``inverse Mercedes Ansatz'',
which is discussed in section\ref{sec5} as the best fit form 
for the gluonic excitation energy $\Delta E_{\rm 3Q}$.
{Form~2} is a simplified form of {Form~1}.
We examine Form~3-5 and Form~6-8 as the typical single-valued function of $L_{\rm min}$ and $L_{\Delta}$, respectively.
As a possibility, the gluonic excitation energy 
$\Delta E_{\rm 3Q}$ may be controlled by the ``typical length" of the static 3Q system, 
and hence we examine also Form~9 and Form~10, regarding 
$\max (x_1+x_2,x_2+x_3,x_3+x_1)$ and $\max (a,b,c)$ 
as the typical length of the Y-type flux-tube system. 

For each of Form 1-10, we show $\chi^2/N_{\rm DF}$ as the fit result 
for the lattice data of $\Delta E_{\rm 3Q}$ 
together with the best-fit parameters 
in Tables~\ref{tab.trials1} and \ref{tab.trials2} at $\beta$=5.8 and 6.0, respectively.

As a result, Form~1, the ``inverse Mercedes Ansatz'', is found to be 
the best fit function for $\Delta E_{\rm 3Q}$, which reproduces the lattice data of $\Delta E_{\rm 3Q}$ with 
the smallest $\chi^2/N_{\rm DF} \sim 1$.

\begin{table}[ht]
\caption{The best-fit analysis for the lattice QCD data of 
the gluonic excitation energy $\Delta E_{\rm 3Q}$ at $\beta$=5.8  
with various trial fit functions. 
For each form, the best-fit parameters $a_i$ are listed with the $\chi^2/N_{\rm DF}$.
\label{tab.trials1}}
\begin{tabular}{p{27pt} p{55pt} p{55pt} p{55pt} c }
\hline\hline
Form & $a_1$ & $a_2$ & $a_3$ & $\chi^2/N_{\rm DF}$ \\ \hline
(A1) &  1.4329(340) &  0.5500(105) &  0.77 & 
%41.2/(24-3)=1.96 \\ \hline
1.96 \\
(A2) &  2.0996(501) &  0.4936(117) &  0.55 & 
%45.4/(24-3)=2.16 \\ \hline
2.16 \\
(A3) &  2.6676(3674) & 1.7127(3568) &  0.4734(339) & 
%72.3/(24-3)=3.44 \\ \hline
3.44 \\
(A4) & -0.2017(2759) & 0.3210(293) &  1.4428(928) & 
%68.6/(24-3)=3.27 \\ \hline
3.27 \\
(A5) &  0.3818(827)  & 1.4193(1612)&  0.1044(1879)& 
%69.0/(24-3)=3.29 \\ \hline
3.29 \\
(A6) &  5.1692(7427) & 3.4586(6932)&  0.4515(369) & 
%45.7/(24-3)=2.18 \\ \hline
2.18 \\
(A7) & -0.0067(5676) & 0.3417(334) &  1.8386(1740)& 
%42.4/(24-3)=2.02 \\ \hline
2.02 \\
(A8) &  0.3269(830)  & 1.8641(1922)& -0.0411(2578)& 
%42.4/(24-3)=2.02 \\ \hline
2.02 \\
(A9) &  0.9341(213)  &  0.6304(89) &  ---         & 
%245.9/(24-2)=11.71 \\ \hline
11.71 \\
(A10) &  0.8065(183)  &  0.6352(87) &  ---         & 
%269.7/(24-2)=12.84 \\ \hline\hline
12.84 \\ \hline \hline
\end{tabular}
\end{table}

\begin{table}[ht]
\caption{
The best-fit analysis for the lattice data of 
$\Delta E_{\rm 3Q}$ at $\beta$=6.0 
with various trial fit functions. 
\label{tab.trials2}}
\begin{tabular}{ p{27pt} p{55pt} p{55pt} p{55pt} c }
\hline\hline
Form & $a_1$ & $a_2$ & $a_3$ & $\chi^2/N_{\rm DF}$ \\ \hline
(A1) &  1.3486(277) &  0.4252(68) &  1.03 & 
%76.8/(73-3)=1.10 \\ \hline
1.10 \\
(A2) &  1.9810(410) &  0.3860(73) &  1.01 & 
%106.7/(73-3)=1.52 \\ \hline
1.52 \\
(A3) &  2.3757(2175)& 2.0816(3065)&  0.3794(146) & 
%181.2/(73-3)=2.59 \\ \hline
2.59 \\
(A4) & -0.5066(2125)& 0.2871(164) &  1.0987(459) & 
%177.3/(73-3)=2.53 \\ \hline
2.53 \\
(A5) &  0.4284(559) & 1.0722(495) &  0.1760(694) & 
%177.8/(73-3)=2.54 \\ \hline
2.54 \\
(A6) &  4.6197(4238)& 4.2849(5709)&  0.3644(156) & 
%110.1/(73-3)=1.57 \\ \hline
1.57 \\
(A7) & -0.3829(4184)& 0.3065(180) &  1.3764(780) & 
%108.0/(73-3)=1.54 \\ \hline
1.54 \\
(A8) &  0.3659(538) & 1.3968(512) &  0.0862(919) & 
%108.0/(73-3)=1.54 \\ \hline
1.54 \\
(A9) &  0.82575(164) &  0.4846(63) &  ---         & 
%607.1/(73-2)=8.67 \\ \hline
8.67 \\
(A10) &  0.7164(143) &  0.4864(63) &  ---         & 
%651.5/(73-2)=9.31 \\ \hline
9.31 \\ \hline \hline
\end{tabular}
\end{table}

\begin{itemize}
\item
Form 1 (The inverse Mercedes Ansatz)
\begin{eqnarray}
\Delta E_{\rm 3Q}
=\frac{a_1}{L_{\rm \overline{Y}}}+a_2 
=\frac{a_1}
{\sum_{i}\sqrt{x_i^2-x_ia_3+a_3^2}}+a_2
\end{eqnarray}
\item
Form 2
\begin{eqnarray}
\Delta E_{\rm 3Q} 
&=&\frac{a_1}
{\sum_{i}\sqrt{x_i^2+a_3^2}}+a_2
\end{eqnarray}
\item
Form 3
\begin{equation}
\Delta E_{\rm 3Q}=\frac{a_1}{L_{\rm min}+a_2}+a_3
\end{equation}
\item
Form 4
\begin{equation}
\Delta E_{\rm 3Q}=\frac{a_3}{(L_{\rm min}+a_1)^{a_2}}
\end{equation}
\item
Form 5
\begin{equation}
\Delta E_{\rm 3Q}=\frac{a_2}{{L_{\rm min}}^{a_1}}+a_3
\end{equation}
\item
Form 6
\begin{equation}
\Delta E_{\rm 3Q}=\frac{a_1}{L_{\Delta}+a_2}+a_3
\end{equation}
\item
Form 7
\begin{equation}
\Delta E_{\rm 3Q}=\frac{a_3}{(L_{\Delta}+a_1)^{a_2}}
\end{equation}
\item
Form 8
\begin{equation}
\Delta E_{\rm 3Q}=\frac{a_2}{{L_{\Delta}}^{a_1}}+a_3
\end{equation}
\item
Form 9
\begin{equation}
\Delta E_{\rm 3Q}=\frac{a_1}{\max (x_1+x_2,x_2+x_3,x_3+x_1)}+a_2
\end{equation}
\item
Form 10
\begin{equation}
\Delta E_{\rm 3Q}=\frac{a_1}{\max (a,b,c)}+a_2
\end{equation}
\end{itemize}

\end{document}